\documentclass[aps,twocolumn,prx,amsfonts,showpacs,superscriptaddress,longbibliography]{revtex4-2}
\usepackage{verbatim,epsfig,amsmath,amssymb,bm,epsf,graphicx,psfrag,bbold,amsthm,amsfonts}
\usepackage[bottom]{footmisc}
\usepackage{hyperref}
\usepackage{cleveref} 
\usepackage{enumitem}
\usepackage{framed}
\usepackage{mathrsfs}
\usepackage{esint}
\setlist{nosep}
\usepackage{placeins}
\usepackage[all]{xy}
\usepackage{color}
\usepackage[utf8]{inputenc}
\usepackage{float}
\usepackage{natbib}
\usepackage{tikz-cd}
\usepackage{verbatim}
\usepackage{leftidx}
\usepackage{physics}
\usepackage{ulem}

\usepackage{pifont}
\newcommand{\cmark}{\ding{51}}%
\newcommand{\xmark}{\ding{55}}%

\newcommand{\bk}{\bm k}
\newcommand{\bp}{\bm p}

\newcommand{\br}{{\bm r}}

\theoremstyle{plain}

\theoremstyle{definition}

\theoremstyle{remark}

\begin{document}
\title{Charge transfer spin-polarons and ferromagnetism in weakly doped AB-stacked TMD heterobilayers}

\author{Daniele Guerci}
\affiliation{Department of Physics, Massachusetts Institute of Technology, Cambridge, MA, USA}
\affiliation{Center for Computational Quantum Physics, Flatiron Institute, New York, New York 10010, USA}

\author{J. H. Pixley}
\affiliation{Department of Physics and Astronomy, Center for Materials Theory, Rutgers University, Piscataway, New Jersey 08854, USA}
\affiliation{Center for Computational Quantum Physics, Flatiron Institute, New York, New York 10010, USA}

\author{Andrew Millis}
\affiliation{Center for Computational Quantum Physics, Flatiron Institute, New York, New York 10010, USA}
\affiliation{Department of Physics, Columbia University, 538 Wst 120th Street, New YOrk, NY 10027}

\date{\today}

	\begin{abstract}
We study the  formation of ferromagnetic and  magnetic polaron states in weakly doped heterobilayer transition metal dichalcogenides in the ``heavy fermion'' limit in which one layer hosts a dense set of local moments and the other hosts a low density of itinerant holes.
We show that interactions among the carriers in the itinerant layer induces a ferromagnetic exchange.
We  characterize the ground state finding a competition, controlled by the carrier concentration and interlayer exchange, between a layer decoupled phase of itinerant carriers in a background of local moments, a fully polarized ferromagnet and a canted antiferromagnet. 
In the canted antiferromagnet phase the combination of the in-plane 120$^{\circ}$ N\'eel order and Ising spin orbit couplings induces winding in the electronic wavefunction giving rise to a topologically non-trivial spin texture and an observable anomalous Hall effect. 
At larger carrier density the ferromagnetically ordered phase transitions into a paramagnetic heavy Fermi liquid state. This theory enables a comprehensive understanding of the existing experimental observations while also making predictions including experimental signatures enabling direct imaging of spin polaron bound states with scanning tunneling microscopy. Our work shows that the prevailing paradigm of the (Doniach) phase diagram of heavy fermion metals is fundamentally modified in the low doping regime of heterobilayer transition metal dichalcogenides.
\end{abstract}
 
\maketitle

{\it Introduction.---}
Stacked and twisted  two-dimensional (2D) van der Waals heterostructures provide a tunable experimental platform in which a wide range of accessible  quantum phases of matter can be ``designed'' by carefully choosing the constituent materials and  their stacking configuration~\cite{andrei2021marvels,kennes2021moire}. The transition metal dichalcogenides (TMDs)~\cite{Mak2022SemiconductorMM} are a versatile platform offering a range of tunability across  moire patterns, while the ability to stack chemically different monolayers enables the realization of strongly correlated phenomena beyond the Hubbard model limit and its destruction by variation of carrier concentration and magnetic field in stacked and selectively gated MoTe$_2$/WSe$_2$ bilayers. These experiments have opened a new paradigm in moire heavy fermion emulators~\cite{Zhao_2023,zhao2024emergence}. 

The ability to continuously control the doping of moir\'e heterobilayers is not easily accessible in solid state compounds, allowing direct access to the physics of electronic doping, including topological states beyond the conventional phase diagram of heavy fermion materials~\cite{Guerci_2023,prb_daniele_2024,Dalal_2021,Kumar_2022,xie2024topologicalkondosemimetalsemulated,mendezvalderrama2024correlatedtopologicalmixedvalenceinsulators}. The recent observation of doping induced ferromagnetism and its competition with the heavy-Fermi liquid at larger filling in MoTe$_2$/WSe$_2$ bilayers~\cite{Zhao_2023,zhao2024emergence} is of particular interest in this context.  A natural interpretation of the experiments is that the 120$^\circ$ antiferromagnetic ordering believed to occur in the undoped (no itinerant carriers) material develops upon doping a non-zero out of plane canting with mean canting angle $\theta$ proportional to the carrier concentration $x$.
This observation suggests that spin polarons may form when carriers are injected above the 120$^\circ$ magnetic background, a question that is intimately connected to the stability of antiferromagnetism~\cite{Coleman_2001,Senthil_2004,Biermann_2005,Yamamoto_2007,Koga_2008} when coupled to a Fermi liquid.

In this work we focus on the low doping regime of MoTe$_2$/WSe$_2$ bilayers at a fillling $n=1+x$ with $x \ll 1$ and provide a theoretical description of the formation of the ferromagnetic insulator phase~\cite{Li_2021}. We employ the model and framework~\cite{Guerci_2023,prb_daniele_2024}  previously used to capture the heavy Fermi liquid regime but, crucially, we include the repulsion between itinerant carriers, neglected in previous thoeretical treatments by us and others~\cite{Guerci_2023,prb_daniele_2024,PhysRevResearch.5.L042033,Seifert}. Differently from the recent theoretical analysis in Refs.~\cite{Seifert,zhao2024emergence} we include also  the Ising spin-orbit coupling~\cite{kormanyos2015k,Zhang_2021}, which lowers the spin symmetry from SU(2) to U(1). 
We show that  interactions among the doped carriers drive virtual fluctuations that drastically modify the local magnetic interactions. At the same time,  this leads to novel and concrete experimental predictions that will open the door to visualize spin-polarons in a triangular lattice 120$^\circ$ antiferromagnetic background.  
We emphasize that these ingredients, which fundamentally go beyond the conventional  paradigm of heavy fermions, play a crucial role in giving rise to ferromagnetism and stabilizing an anomalous Hall effect across the phase diagram.

Our theory provides several concrete experimental predictions, including a stable canted antiferromagnetic phase in the presence of very weak doping, in good agreement with the antiferromagnetic Curie-Weiss temperature measured experimentally~\cite{zhao2024emergence} as well as the quantized anomalous Hall state at $x=0$ by self-doping the charge transfer band for small values of $\Delta$~\cite{Li_2021,tao2022valleycoherent,Devakul_2021,KTLaw_2022,Dong_2023}. Furthermore, we show that in this regime spin-polaron states can form due to coupling dilute electronic carriers to the triangular N\'eel antiferromangetic spin textures. Upon increasing doping further, the itinerant electrons destabilize the canted antiferromagnetic state, leading to a ferromagnetic state in which all spins (in both layers) contribute to the order. In addition to giving rise to ferromagnetism, these interactions also produce an $s$-wave component of the Kondo coupling, which is ferromagnetic and hence irrelevant (in the  renormalization group sense~\cite{Anderson_1970}). We show that at the single impurity limit, the $p$-wave hybridization competes with the local moment ferromagnetic state eventually inducing a heavy Fermi liquid at large enough doping whose scale is crucially affected by various relevant perturbations discussed at the end of the manuscript. Incorporating disorder within each layer at the level of symmetry then provides an elegant description of the insulating metallic states observed experimentally~\cite{zhao2024emergence}.

{\it Model.---} We consider a model on the moire lattice formed from the lattice mismatch of WSe$_2$ and MoTe$_2$, which yields a moir\'e lattice constant $a_M = a/\delta \approx 5$nm~\cite{Zhang_2021} and we consider hole dopings $n=1+x$ with $0\leq x <1$ so that we need to retain only the highest-lying hole bands in the WSe$_2$ and MoTe$_2$ layers. The key features of the heterobilayer are described by a  Hamiltonian that contains within-layer and intra-layer hoppings and, crucially, local respulsive interactions on each layer: 
\begin{equation}
    \begin{split}
    H=&- t_{\rm W}\hspace{-.3cm}\sum_{\langle \br,\br'\rangle\in \rm W} e^{-i\nu_{\br,\br'}\varphi_{SO} s_\sigma} c^\dagger_{\br\sigma}c_{\br'\sigma}
    -t_{\rm M}\hspace{-.3cm}\sum_{\langle\br,\br'\rangle\in{\rm M}}f^\dagger_{\br \sigma} f_{\br'\sigma}
    \\
    &-t_\perp\sum_{\br\in {\rm M}}\sum_{j=1}^{3}(f^\dagger_{\br\sigma} c_{\br+\bm u_j\sigma}+h.c.)  +\sum_{\ell}U_{\ell}\sum_{\br \in {\ell}} n_{\br \uparrow} n_{\br \downarrow} \\
    & -\Delta(N_{\rm M}-N_{\rm W})/2
    \end{split}
    \label{Hamiltoninan_lattice}
\end{equation}
with $(t_{\ell },U_{\ell})$ the intralayer hopping amplitude and on-site interaction for $\ell=$W and $\ell=$Mo layers, respectively, $t_\perp$ interlayer hopping and $\Delta$ energy off-set tunable by displacement field. The Ising spin-orbit coupling induces a hopping phase $\varphi_{SO}$ which can be represented in different ways. We choose a gauge in which the top of the Mo-layer valence band is at the $\gamma$ point and the phase occurs explicitly in the $W$ layer hopping taking the value $\varphi_{SO}=2\pi/3$ in MoTe$_2$/WSe$_2$.
In Eq.~\eqref{Hamiltoninan_lattice} $\nu_{\br,\br'}=\pm 1$ defines the direction of hopping and the spin orbit coupling is manifest in the dependence of the hopping phase on the sign of the $z$ component of the  spin $s_\sigma$. We explore the regime $U_{\rm M}> U_{\rm W}\gg \Delta$ and $\Delta>t_\perp$ 
where at filling factor $\nu=1$ the holes reside in the Mo layer and forms a Mott insulator with 120$^\circ$ ordering. Henceforth, we refer to the electrons on the Mo layer as $f$ electrons and electrons on the W layer as c electrons.

\begin{figure}
    \centering
\includegraphics[width=\linewidth]{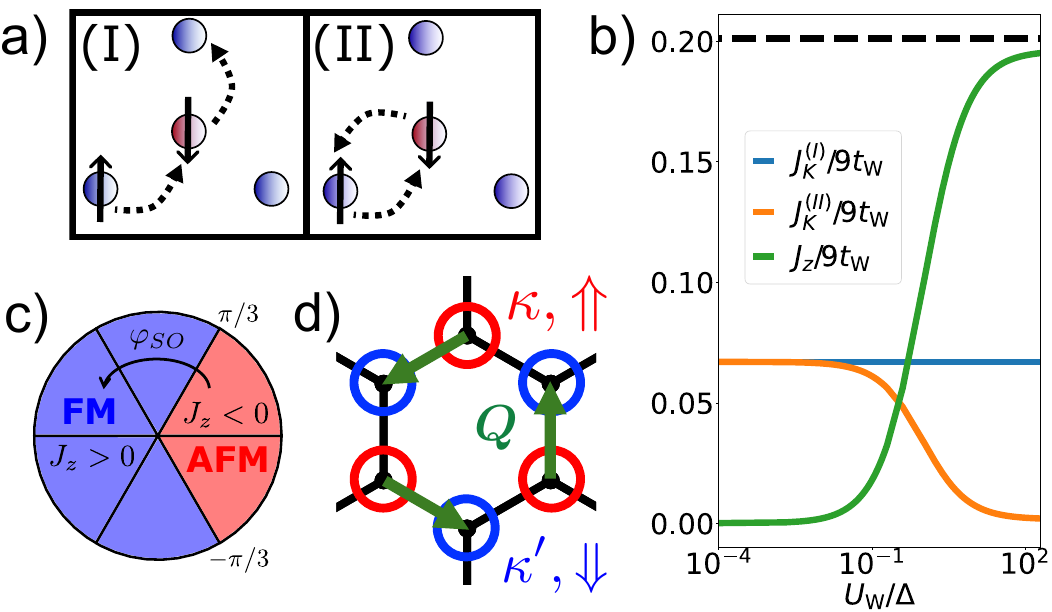}
    \caption{{\bf Spin orbit coupling induced ferromagnetism}: Panel a) shows the double exchange $J^{(I)}_K$ and superexchange processes $J^{(II)}_K$. Panel b) shows the evolution of the couplings $J^{(I)}_{K}$, $J^{(II)}_{K}$ and $J_z=3[J^{(I)}_{K}-J^{(II)}_{K}]$ as a function of $U_{\rm W}/\Delta$ at $U_M/\Delta=33$. Dashed line shows the asymptotic value $3J^{(I)}_K$. Panel c) shows the sign of $J_z$ obtained by projecting $J_{\bk,\bp}$ on the Fermi surface as a function of the Ising spin-orbit coupling phase $\varphi_{SO}$ represented as the polar angle; radial coordinate has no meaning. d) Sketch of the spin flip scattering process $\Psi^\dagger_{\bk\Uparrow}\Psi_{\bk\Downarrow}$ induced by the coupling to $S^-$ spin-valley locking introduces the momentum selection rule $\bm Q=\kappa'-\kappa$.
    }
    \label{fig:exchange_couplings}
\end{figure}

{\it Magnetic Correlations.---} 
The origin of the ferromagnetic correlations can be understood by taking the large $U_{\rm M}$ of the Hamiltonian. Assuming $\langle n_f\rangle\approx 1$ the central site becomes a spin degree of freedom exchange coupled to the itinerant carriers. Including charge fluctuations within second order perturbation theory we find the Kondo exchange (as depicted in Fig.~\ref{fig:exchange_couplings}): 
\begin{equation}\label{Kondo_exchange}
H_K =\sum_{\br\in \text{Mo}}\frac{\hat{\bm S}_\br}{2}\cdot\sum_{i,j}\left(J_K^{(I)}+J_K^{(II)}\delta_{ij}\right) c^\dagger_{\br +\bm u_i} \bm \sigma c_{\br +\bm u_j}
\end{equation}
where $\bm u_j$ is the vector connecting the two triangular lattices shown in Fig.~\ref{fig:exchange_couplings} (a),  and $J^{(I)}_K=2t^2_\perp U_{\rm M}/[\Delta(U_{\rm M}-\Delta)]\rightarrow 2t^2_\perp/\Delta$ and $J^{(II)}_K=2t^2_\perp (U_{\rm M}+U_{\rm W})/[(U_{\rm M}-\Delta)(U_{\rm W}+\Delta)]\rightarrow 2t^2_\perp/(U_{\rm W}+\Delta)$~\cite{Seifert} in the $U_{\rm M}\rightarrow\infty$ limit, originating from the second order processes in Fig.~\ref{fig:exchange_couplings}a). 
The second expression is the limit of very large $U_{\rm M}$. The repulsion $U_{{\rm W}}$ changes the relative amplitude of the two exchange processes between the itinerant electrons and the local moments shown in Fig.~\ref{fig:exchange_couplings}a). The interplay between these two effects can be easily seen in momentum space where the non-local structure of the exchange~\eqref{Kondo_exchange}  leads in the plane-wave basis to a $\bk$-dependent coupling $H_K=\sum_{\bk,\bp}J_{\bk,\bp}\hat{\bm S}_{\bp-\bk}\cdot c^\dagger_{\bk}\bm \sigma c_{\bp}/2$ with
\begin{equation}
    J_{\bk,\bp} = J^{(I)}_KV^*_{\bk}V_{\bp} +  \left(J^{(II)}_K-J^{(I)}_K\right) V_{\bp-\bk}
\end{equation}
where $V_{\bk}=\sum_{j=1}^3 e^{i\bk\cdot\bm u_j}$. At low doping $x\ll1$ carriers of the itinerant layer are located around $\kappa$ (spin up) and $\kappa^\prime$ (spin down) of the mini Brillouin zone. Because $V_\gamma=3$ while $V_\kappa=V_{\kappa^\prime}=V_{\kappa-\kappa^\prime}=0$, and $J_K^{(I)}>J_K^{(II)}$ the exchange coupling is negative (ferromagnetic) and only nonzero for the $z$ component ($J_{\kappa,\kappa'}=0$), leading to $J_{\kappa,\kappa}=J_{\kappa',\kappa'}\equiv -J_z=-6t^2_\perp U_{\rm W}/\left[\Delta(U_{\rm W}+\Delta)\right]<0$. Additionally, for small momentum deviations around the high-symmetry points $\kappa/\kappa'$ the spin exchange acquires the characteristic $p$-wave form leading eventually at large enough doping $x$ to the topological heavy Fermi liquid~\cite{Guerci_2023,prb_daniele_2024}.
The perturbative result is confirmed by exact diagonalization on the ``tetrahedron'' model composed by four sites, two carrier and given in the SM~\cite{supplementary}. Specifically, we find that ferromagnetic correlations develop in the local moment region $n_f=1$ only for finite repulsion between itinerant carriers $U_{\rm W}\neq 0$ consistently with Fig.~\ref{fig:exchange_couplings}b).  

Before concluding, we observe that the value of the  Ising spin orbit coupling induced hopping phase $\varphi_{SO}=2\pi/3$ plays a crucial role in our considerations; this value is a consequence of the AB stacking structure. Generalizing to other stackings gives different values of $\varphi_{SO}$ which lead to different magnetic correlations as sketched in Fig.~\ref{fig:exchange_couplings}c). A ferromagnetic coupling is realized in the range $\varphi_{SO}\in(\pi/3,5\pi/3)$ while an antiferromagnetic coupling is found in the interval $\varphi_{SO}\in(-\pi/3,\pi/3)$, which also confirmed by numerical diagonalization of the tetrahedron model (see SM~\cite{supplementary}). Interestingly the two configurations $\varphi_{SO}=\pi/3$  ($\varphi_{SO}=-\pi/3$) where the itinerant carriers display Fermi pockets around $\gamma$ and $\kappa$ ($\kappa'$) lead to a unique kind of Kondo frustration, where one part of the Fermi surface experiences a ferromagnetic coupling to the local moments while the other experiences an antiferromagnetic one.     

\begin{figure}
    \centering
    \includegraphics[width=\linewidth]{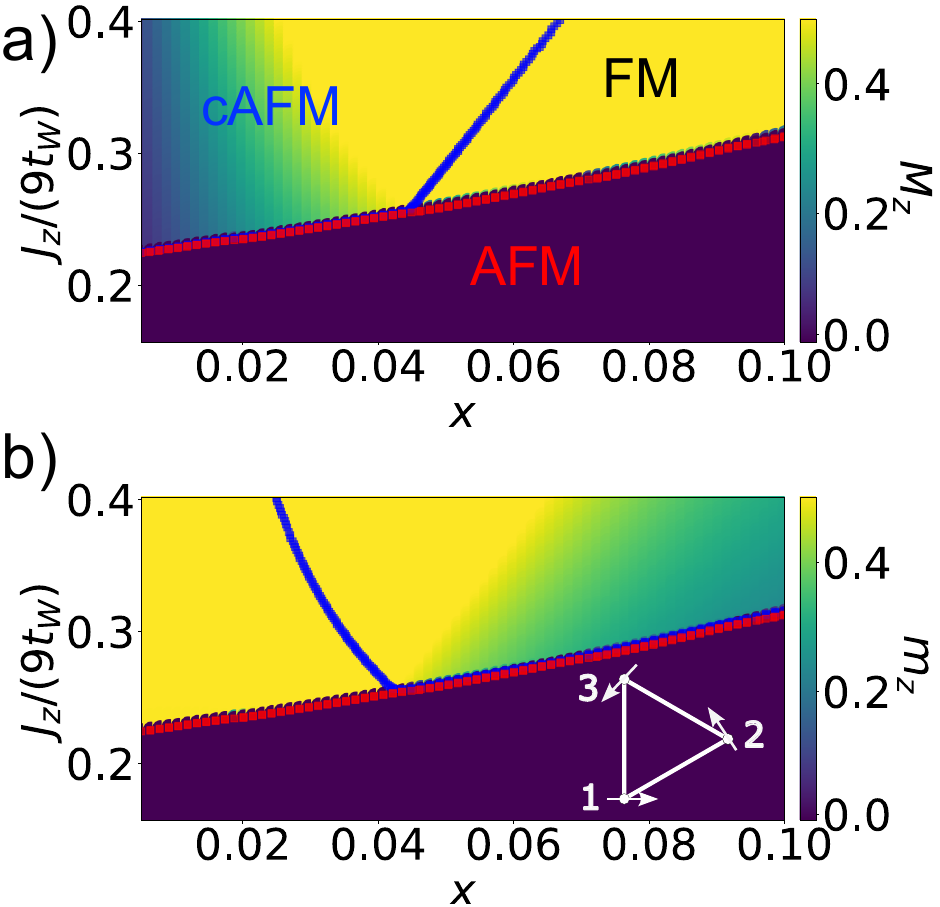}
    \caption{{\bf Hartree-Fock Phase Diagram}: Panel (a): color plot of magnitude of out of plane magnetization $M_z$ of Mo-plane local moments as function of doping $x$ and $J_z/(9t_{\rm W})$ with $9t_{\rm W}$ bandwidth of the itinerant carriers. 120$^\circ$ planar antiferromagnet (AFM), canted 120$^\circ$  antiferromagnet (cAFM) and fully polarized ferromagnetic (FM) phases are labelled. The red line indicates a first order transition across which $M_z$ changes discontinuously; the blue line indicates the onset of full polarization $M_z=0.5$. 
    Panel (b): color plot of magnitude of out of plane magnetization $m_z=(n_\uparrow-n_\downarrow)/2$ (in units of itinerant electron density $x$) of W plane itinerant electrons  as function of doping $x$ and $J_z/(9t_{\rm W})$ with $9t_{\rm W}$ bandwidth of the itinerant carriers.  Red and blue lines as in panel (a). The spin configuration in a triangular plaquette with helicity $\chi=+$ is also sketched, notice that the out-of-plane component in the cAFM gives rise to spin chirality $\chi_{\triangle}=\bm S_1\cdot(\bm S_2\times\bm S_3)$ and anomalous Hall effect. Calculations are performed setting $t_{\rm W}=6$meV, $t_\perp=2$meV, $J_H=0.1$meV, $U_{\rm W}=70$meV, $U_{\rm M}=100$meV.}
    \label{fig:phase_diagram_magnetization}
\end{figure}

{\it Ferromagnetic correlations from doping and anomalous Hall Fermi liquid.---} The classical energy of doped WSe$_2$ in the background of the local moments of MoTe$_2$ can be readily obtained starting from the vacuum corresponding to the local moment configuration where $\langle n_f\rangle\approx 1$ and the local moments interact via the Heisenberg exchange $J_H\sum_{\langle\br,\br'\rangle}\bm S_{\br}\cdot\bm S_{\br'}$ with $J_H=4t^2_{\rm M}/U_{\rm M}+4t_{\rm M}t^2_\perp/\Delta^2$ that in the following will be treated as a tunable parameter.  
The classical ground state for $\Delta$ large enough reads $\ket{\Phi}=\prod_{\bm r\in {\rm M}}U_{\br}{[f^\dagger_{\br\uparrow}+f^\dagger_{\br\downarrow}]}/{\sqrt{2}}U^\dagger_{\br}\ket{0}$ where $U_{\br}=\exp\left[-iS^z_{\br}(\bm Q\cdot\br+\phi)\right]$ is a space-dependent $U(1)$ rotation around the $z$ axis, $S^z_{\br}=[f^\dagger_{\br\uparrow}f_{\br\uparrow}- f^\dagger_{\br\downarrow}f_{\br\downarrow}]/2$ and $f_{\br}$ are the local moments in the MoTe$_2$ layer. 
The resulting ordered spin texture describes an intervalley coherent  120$^\circ$ magnetically ordered state with an order parameter $\mel{\Phi}{\bm S_{\br}}{\Phi}=[\cos(\bm Q_{\rm or}\cdot\br+\phi),\sin(\bm Q_{\rm or}\cdot\br+\phi),0]/2$ with modulation on the moir\'e scale $\bm Q_{\mathrm{or}}=\pm \bm Q=\pm \left(\bm \kappa'-\bm \kappa\right)$ is the N\'eel vector and the orientation of the triangular order is set by $\phi$.
The 120$^\circ$ order breaks the time-reversal symmetry $\mathcal T=i\sigma^y\mathcal K$ with $\mathcal K$ complex conjugation as well as the U(1) valley symmetry generated by $S^z$. 
Combining the broken generators $\mathcal T$ with $\sigma^z$ we get $\Theta=\sigma^z\mathcal T=\sigma^x\mathcal K$ symmetry of the 120$^\circ$ ordered phase. 
The magnetic state acquires a helicity $\chi=\bm z \cdot\sum_{jkl} \varepsilon_{jkl} \left(\bm S_k\times \bm S_l\right)/(3\sqrt{3})=\pm 1$ with the $\bm S_{1,2,3}$ spins on the triangle shown in the inset of Fig.~\ref{fig:phase_diagram_magnetization}b), where  the sign of the helicity is determined by the sign of $\bm Q$. In an isolated MoTe$_2$ layer the two helicity states are degenerate. 
The combination of inversion symmetry breaking in the heterobilayer and spin orbit coupling favors one of the two helicity states. The specifics depend on the bilayer geometry and other conventions as well as the doping.  The symmetry breaking can be seen the present model as follows: the ordering wave vector $\bm Q_{\mathrm{or}}=+\bm Q$ scatters an electron in the WSe$_2$ layer from the $\kappa^\prime$ (spin down) pocket to the $\kappa$ (spin up) pocket,  thus only coupling to the spin operator $\sigma^+$ shown in Fig.\ref{fig:exchange_couplings}d).  This means that added carriers break the mirror symmetry $\mathcal{M}_x$  of the undoped state, favoring the $\chi=+1$  chirality. In contrast, doping the $\bm Q_{\mathrm{or}}=-\bm Q$ configuration with $\chi=-1$ chirality cannot couple to the spin flip  and is higher in energy as a consequence. For this reason, in our model doping selects the $\chi=+1$ chirality.

We now consider the evolution of the classical ground state upon doping. For the combination of $\Delta$ and $U_{\rm M}$ considered here, the added carriers go into the $\rm W$ plane. Appendix ~\ref{app:classical_limit} presents a detailed derivation of the full Hamiltonian for the doped materials.  Here for ease of discussion we consider  the Hartree-Fock energy per site,  given by
\begin{equation}\label{classical_energy}
    \begin{split}
         &\mathcal E_{\rm cl}[\{\bm M,\bm m\}]=-3 J_HM^2_\parallel/2+3J_ H M^2_z-J_zm_zM_z\\
         &+\int_{\bk}\frac{\bk^2\rho_{\bk}}{2m^*_{\rm W}}+J_\perp M_z \bm k^2s^z_{\bk} +J_\perp M_\parallel\frac{k_-^2s^+_{\bk}+k_+^2 s^-_{\bk}}{2},
    \end{split}
 \end{equation}
where $M_\parallel$ and $M_\perp$ are respectively the magnitudes of the in-plane and out of plane components of the local moment order parameter,  $\rho_{\bk}=\langle\Psi^\dagger_{\bk}\Psi_{\bk} \rangle$, $\bm s_{\bk}=\langle \Psi^\dagger_{\bk}\bm\sigma\Psi_{\bk}\rangle/2$,  $\int_{\bk}\cdots=\int^{\Lambda}d^2\bk\cdots/\Omega_{\rm BZ}$ the integral is performed on a small region of radius $\Lambda$ of the entire mini Brillouin zone (mBZ), $\bm m$ is the magnetization of the itinerant electrons and we have introduced $\Psi_{\bk\Uparrow(\Downarrow)}\equiv c_{\kappa(\kappa')+\bk\uparrow(\downarrow)}$ with $|\bk|\le\Lambda$. 
The  last ($J_\perp M_\parallel$) term in Eq.~\eqref{classical_energy} is present in the energy only for the $+1$ chirality, for the reasons given above; the energy gain from this term is responsible for breaking the degeneracy between the two 120$^\circ$ ordering with helicity $\chi=\pm$ states degenerate in the undoped limit.
We emphasize that the small momentum result in Eq.~\eqref{classical_energy} is the most general result according to the symmetries $C_{3z}$, $\mathcal M_y$, and $\Theta$, i.e. $k_+\sigma^++h.c.$ is consistent with $\mathcal M_y$ and $C_{3z}$ but is not allowed by $\Theta$. 

%
%
$\mathcal E_{\rm cl}$~\eqref{classical_energy}  is characterized by different energy scales $J_H$ which favours an in-plane $120^\circ$ degree order, $J_z=3\left[J^{(I)}_{K}(1-x)^2-J^{(II)}_K\right]$ and $J_\perp=3J^{(I)}_K(1-x)^2/4$ describing the coupling of the local moments to the itinerant carriers. 
The classical energy does not depend on $\phi$ defining the in-plane orientation of the $120^\circ$ order which is set $\phi=0$ for convenience. 

The phase diagram in Fig.~\ref{fig:phase_diagram_magnetization}a) and~\ref{fig:phase_diagram_magnetization}b) is obtained by minimizing the Hartree-Fock energy $\mathcal E_{\rm cl}$~\eqref{classical_energy} and shows the evolution of the out of plane magnetization $M_z$ and $m_z$ as a function of $J_z$ and of the doping $x$, respectively. 
Above a critical value of $J_z$ highlighted by the red dotted line in Fig.~\ref{fig:phase_diagram_magnetization}a) and b) we find that the ground state acquires a nonzero out of plane magnetization and the ground state breaks the emergent symmetry $\Theta$ of the 120$^\circ$ N\'eel phase (AFM). 
In this regime there are two different magnetic ordering separated by the blue dots in Fig.~\ref{fig:phase_diagram_magnetization}a) corresponding to a canted 120$^{\circ}$ N\'eel order (cAFM) and an out of plane ferromagnet (FM). We observe that the energy gain in developing an out of plane magnetization $M_z$ goes like $-J_z m_z M_z$ and scales linearly with the density of carriers $x$ in the itinerant WSe$_2$ layer. Thus, for small $x$, the magnetic energy is parametrically larger than the kinetic energy cost which goes like $\sim x^2$. As a result, for $x\le 0.05$ the itinerant carriers are fully polarized $m_z/x\approx 0.5$ in the region delimitated by the blue line in Fig.~\ref{fig:phase_diagram_magnetization}b). Furthermore,  the magnetization of the local moments $M_z$ evolves continuously as a function of $x$ for a given $J_z$ across the second order transition line highlighted in blue in Fig.~\ref{fig:phase_diagram_magnetization}a).
Conversely, the red line separating the AFM phase from the cAFM and FM phases in Fig.~\ref{fig:phase_diagram_magnetization}a) and b) indicates a pronounced first-order transition, where the out-of-plane magnetization of both the itinerant carriers and the local moments acquires a net ferromagnetic component.

Notably, the region at small filling factor $x$ realizes an anomalous Hall metal.
A finite out of plane magnetization $M_z$ breaks spontaneously the $\Theta$ symmetry, $\Theta \Psi_{\bk\alpha}\Theta^{-1}=\sigma^x_{\alpha\beta}\Psi_{-\bk\beta}$, and in the cAFM region leads to a finite scalar spin chirality $\chi_{\triangle}=\bm S_{1}\cdot(\bm S_2\times\bm S_{3})$~\cite{Guerci_2023,PhysRevResearch.5.L042033} with $\bm S_j$ spin on the triangular plaquette displayed in Fig.~\ref{fig:phase_diagram_magnetization}b). Remarkably, the net spin chirality results in anomalous Hall transport properties of the low-density Fermi liquid~\cite{supplementary,Haldane_2004,Binz_2008,verma2022unified,IvarMartin_20108}, as we demonstrate in the suppplemental.

Our continuum theory also accounts for the $x=0$ transition  to an anomalous Hall state  in good agreement with experiments~\cite{Li_2021,tao2022valleycoherent}, as the charge transfer energy $\Delta$ is varied in Eq.~\eqref{Hamiltoninan_lattice}. 
This regime is accurately described by an extended periodic Anderson model presented recently in Ref.~\cite{prb_daniele_2024} where in addition to the low-energy spin of the Mo layer also holon charge fluctuations are included $f_\sigma=b^\dagger \psi_\sigma$. The continuum theory for the bosonic excitations reads 
\begin{equation}
    \mathcal L_b = \int_{k} \bar b_{k}\left[i\nu -r-\Sigma_b(k)\right]b_k
\end{equation}
where $r\approx\Delta+\mu$ and $\mu$ the chemical potential of the itinerant carriers and $\Sigma_b(k)$ the self-energy correction from interlayer tunneling processes. At a critical $\Delta/t_\perp$ the interlayer charge transfer excitation $b$ becomes critical $r_{\rm eff}=r+\Sigma_b(k=0)=0$ and the itinerant layer is self-doped by the local moments introducing holons $\langle b^\dagger b\rangle >0$ in the lower Hubbard band~\cite{Guerci_2023}. The interlayer ferromagnetic exchange then opens a full topological gap, resulting in a quantum anomalous Hall cAFM state in agreement with detailed lattice calculations~\cite{Devakul_2021}.

\begin{figure}[]
    \centering
\includegraphics[width=0.8\linewidth]{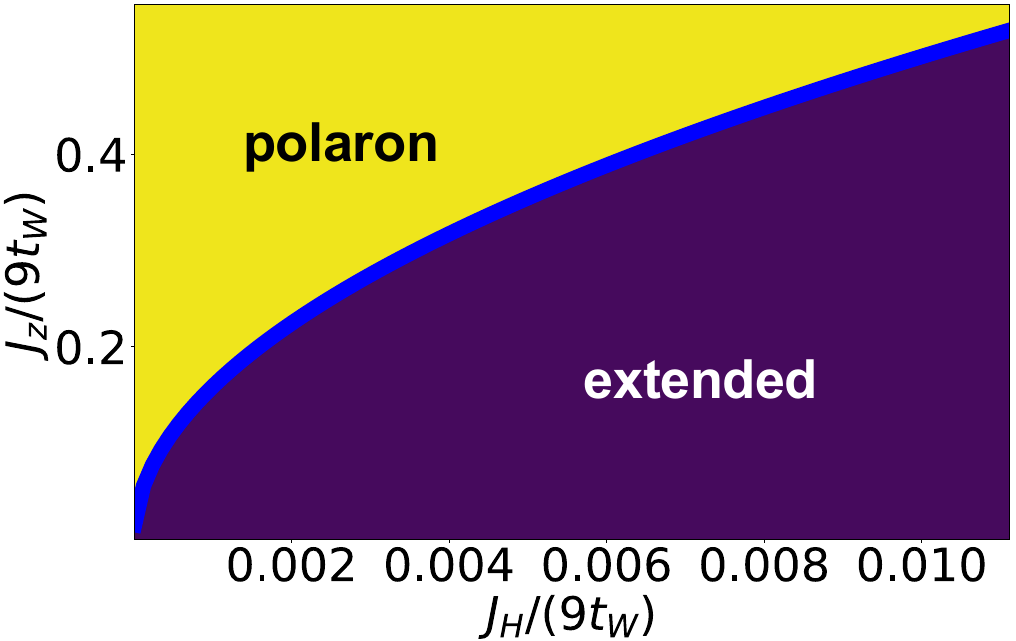}
    \caption{{\bf Polaron Phase Diagram}: Variational phase diagram of the lowest energy magnetic polaron for $J_\perp=0$. The bound state occurs in the blue region of the color map above the critical line, while for smaller values of $J_z$ creating a local out-of-plane canting of the magnetic texture becomes energetically unfavorable. Calculations are performed setting $t_{\rm W}=6$meV. }
    \label{fig:bound_state}
\end{figure}
{\it Magnetic polaron in the continuum limit.---}
The ability of a small density of itinerant carriers to produce a small magnetization raises the question whether a single carrier will produce a magnetic polaron.  To address this possibility we study the low-doping configuration considering the motion of a single carrier in WSe$_2$ allowing for non-uniform distortions of the magnetic texture, following similar resoning developed in Refs.~\cite{PhysRevLett.61.467,Shastry_1981,Siggia_1988,Sigrist_1991}. 
The total low-energy action reads; $\mathcal L =\mathcal L_{s}+\mathcal L_{\psi}+ \mathcal L_c.$
In the absence of doped carriers, fluctuations of the spin texture are described by the nonlinear sigma model (NL$\sigma$M) given in Ref.~\cite{read1989,Nikolic2020,Tchernyshyov_2024,supplementary} describing the coupled fluctuations of the canting angle $\theta$ and the in-plane twist $\phi$. 

The corresponding Lagrangian is $
     \mathcal L_{S}=\int_{\br} [-\theta\partial_\tau\phi+\frac{3J_H}{8}\theta^2+ \frac{9J_H}{32}\left(\nabla_{\br}\phi\right)^2].$ The Lagrangian of the itinerant carriers (wave function $\psi$) in the mean field antiferromagnetic background is  is \begin{equation}\label{fermionic_lagrangian}
     \mathcal L_{\psi}=\int_{\br} \bm\psi^\dagger\left[\partial_\tau+\frac{\bk^2}{2m^*_{W}}+\frac{J_\perp\hat k^2_-}{4}\sigma^++h.c.\right]\bm \psi,
 \end{equation}
where $\int_{\br}\cdots=\int d^2\br/\Omega\cdots$, $\bm \psi=[\psi_\Uparrow,\psi_\Downarrow]^T$. As a result of the U(1) spin symmetry of the theory~\eqref{Hamiltoninan_lattice}, the motion of itinerant carriers couples to the canting angle $\theta$ and the in plane $\phi$ twist around $z$ of the orientation of the three sublattice 120$^\circ$ order leading to the coupling Lagrangian
 \begin{equation}
    \begin{split}\label{L_coupling}
     \mathcal L_{c}
     =\int_{\br}\,  \bm\psi^\dagger\left[ \frac{-J_z}{4}\sigma^z  \theta 
    +\frac{J_\perp}{2\sqrt{3}}
    (\sigma^x\,\partial_y\phi+
    \sigma^y\partial_x\phi)\right]\bm\psi.
    \end{split}
 \end{equation}
  As expected, the Goldstone mode $\phi$ couples to the itinerant carriers via a gradient coupling that vanishes in the limit of zero transfer momentum $\bm q\to0$~\cite{Watanabe_2014,Watanabe_2014_Skx,Watanabe_2020}. On the other hand, the coupling $J_z$ with the net out of plane magnetization $\theta$ competes with the exchange energy $3J_H\theta^2/8$ which forces the spins to have a net vanishing magnetization per triangular plaquette~\cite{read1989,Nikolic2020,Tchernyshyov_2024}. Last, we also observe that the in-plane ferromagnet realized in AA-stacked heterobilayers~\cite{Crepel_2023} has the same effective action $\mathcal L$ but with $J_\perp=0$ in $\mathcal L_\psi$~\eqref{fermionic_lagrangian}. 

For a first investigation of the spin polaron, we consider the saddle point of the action following from this Lagrangian. In the static limit, $\delta\mathcal L/\delta\theta=0$ implies $\theta=J_z\bm\psi^\dagger\sigma^z\bm\psi/3J_H$, which is thus self-consistently determined by the electronic spin magnetization. Introducing the field $\bm e=(-\partial_x\phi,\partial_y\phi)$ describing a local twist of the magnetization texture we find $\delta \mathcal L/\delta e_a=0$ leading to the relation $e_a=8J_\perp\varepsilon_{abz}\bm\psi^\dagger\sigma^b\bm\psi/(9\sqrt{3}J_H)$ with $\varepsilon$ Levi-Civita tensor. Finally, we have the Schr\"odinger equation for $\bm\psi$ obtained by minimizing $\delta\mathcal L/\delta\psi^\dagger_\sigma=0$:
\begin{equation}
    \begin{split}
    \label{saddle_point_action}
     &\left[\frac{\bk^2}{2m^*_{\rm W}}+\frac{J_\perp \hat k^2_-\sigma^++h.c}{4}-\frac{J^2_z\bm\psi^\dagger\sigma^z\bm\psi}{12J_H}\sigma^z\right.\\
     &\left.+\frac{J_\perp(\bm\sigma\times\bm e)_z}{2\sqrt{3}}\right]\bm\psi= \lambda \bm\psi.
    \end{split}
\end{equation}
In the following we focus on the solution of the non-linear equations with $J_\perp=0$ where, at self-consistency, the bound state forms only an out-of-plane spin texture with $\bm e=0$. In this limit, we analytically find that the critical value $J^c_z$ for the formation of a bound state solution is $J^c_z=12\sqrt{\pi t_{\rm W}J_H/2}$ depicted in Fig.~\ref{fig:bound_state}, corresponding to $J^c_z/(9t_W)\approx 0.216$ for $J_H=0.1$meV---slightly below the critical $J^c_z(x\to0)/(9t_{\rm W})\approx0.22$ in Fig.~\ref{fig:phase_diagram_magnetization} for an uniform magnetization.  
We thus suggest that at very low dopings, spin polarons in an antiferromagnetic background provide the most appropriate description of the physics. 
The solution of the bound state problem in the realistic regime $J_\perp\neq 0$ and in the presence of random potentials leading to electron pinning is left to future investigations. Finally, we highlight that the formation of spin polaron can also emerge at $x=0$ from the self-doping of the lower Hubbard band.

\begin{figure}
    \centering
    \includegraphics[width=0.8\linewidth]{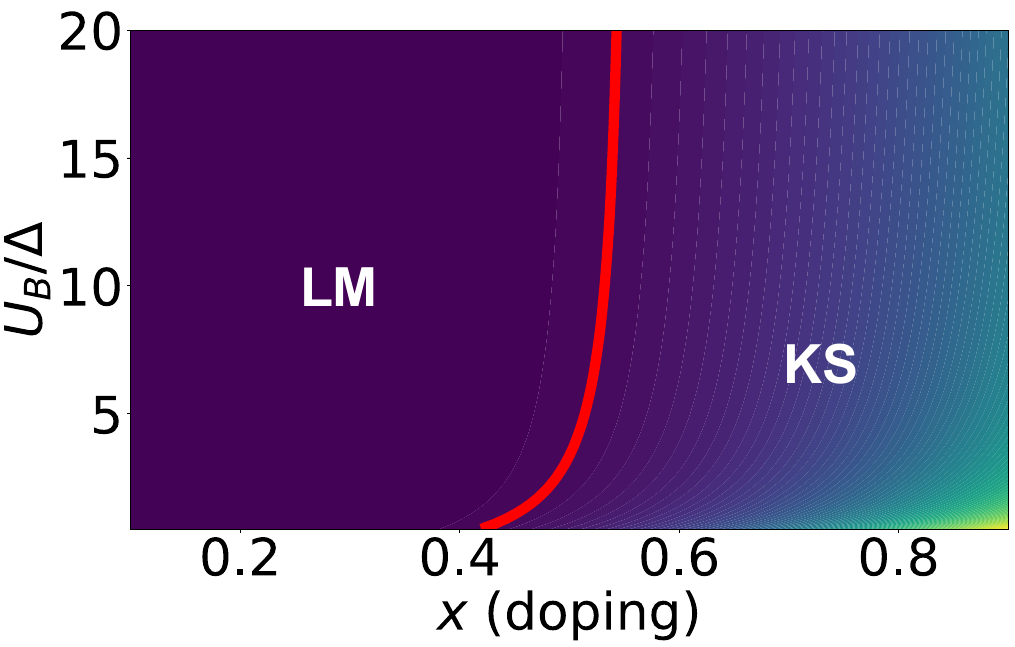}
    \caption{{\bf $p$-wave Kondo screening  in the single impurity limit}: At low-doping, the spin is in the local moment regime. Only above a critical filling $x$ we find Kondo screening. The phase diagram is obtained setting an infrared cutoff to $0.6$K and the color code shows $T_K$. }\label{fig:scaling_law_diagram}
\end{figure}


{\it Screening of the magnetic moments: Poor man's Anderson scaling laws.---} 
Up to this point we have considered the low doping limit, where magnetism is the dominant effect. At higher doping,  Kondo screening is known experimentally to lead to the formation of a heavy fermi liquid, which we now turn to ~\cite{Zhao_2023}. 
Magnetic ordering competes with the Kondo effect~\cite{Doniach_1977} where the screening of the local moments leads to a paramagnetic Fermi liquid state of quasiparticles with large effective mass~\cite{Guerci_2023,prb_daniele_2024}. To describe this competition in the current setting, 
we focus on the single impurity limit of the Kondo interaction in Eq.~\eqref{Kondo_exchange}. Taking advantage of the spherical symmetry of the theory we expand in partial waves $\Psi_{\bk\sigma}=\sum_{m}e^{im\theta_{\bk}}\chi_{m\sigma}(k)$ which due to the three-fold rotational symmetry $C_{3z}$ implies that only couplings between Cooper pairs with equal total angular momentum $J^z=L^z+2S^z$ module 3 are allowed. The resulting exchange is obtained by expanding $H_K$~\eqref{Kondo_exchange} close to the bottom of the itinerant band: 
\begin{equation}\label{Kondo_hamiltonian}
    \begin{split}
    \mathcal H_{K} &= -\frac{J_z-\delta J_z}{2} S^z\bm \chi^\dagger_{0}\sigma^z\bm \chi_0\\
    &-\frac{J_z}{24}S^z \left(\chi^\dagger_{+\Uparrow}\chi_{+\Uparrow}-\chi^\dagger_{-\Downarrow}\chi_{-\Downarrow}\right)\\
    &+\frac{J_\perp-J_z/12}{2}\left(S^+\chi^\dagger_{+\Downarrow}\chi_{-\Uparrow}+h.c.\right)\\
    &+\frac{J_\perp-J_z/12}{2} S^z \left(\chi^\dagger_{-\Uparrow}\chi_{-\Uparrow}-\chi^\dagger_{+\Downarrow}\chi_{+\Downarrow}\right)\\
    &+\frac{J_z}{4\sqrt{3}} \left( S^+\chi^\dagger_{-\Downarrow}\chi_{0\Uparrow}+S^+\chi^\dagger_{0\Downarrow}\chi_{+\Uparrow}+h.c.\right),
    \end{split}
\end{equation}
where we have introduced
\begin{equation}
    \bm \chi_0=\frac{1}{\sqrt{N}}\sum_{\bk}^\Lambda \bm \Psi_{\bk},\quad  \bm \chi_{\pm }=\frac{1}{\sqrt{N}}\sum_{\bk}^\Lambda a k_{\mp}\bm \Psi_{\bk}.
\end{equation}
In the effective Hamiltonian~\eqref{Kondo_hamiltonian} the $s$-wave channel $\bm\chi_0=[\chi_{0\uparrow},\chi_{0\downarrow}]^T$ couples with $[\chi_{+\uparrow},\chi_{-\downarrow}]^{T}$ while the $p$-wave component $[\chi_{-\uparrow},\chi_{+\downarrow}]^T$ with $J_z=L_z+2S_z=0$ does not couple to the other channels. The latter is characterized by an antiferromagnetic interaction and, eventually, leads to screening of the local moment~\cite{Anderson_1970}:
\begin{equation}
    \frac{dJ_p}{d\log s}=m^*_{\rm W}x J^2_p,
\end{equation}
with $J_p(0)=J_\perp-J_z/12$ and $s=T'/T>1$. As already noted in Ref.~\cite{Guerci_2023}, the Kondo temperature from the $p$-wave channel scales as $T_K=\Lambda e^{-1/[m^*_W x J_p(0)]}$ and is exponentially suppressed in the low-doping regime, allowing for a local moment (LM) phase to develop as shown in Fig.~\ref{fig:scaling_law_diagram}. 
While this theory underestimates $T_K$ at low doping, as  Kondo temperature observed experimentally is much larger being on the order of $\sim10$K for $x=0.15$~\cite{Zhao_2023}. This  suggests a key role played by relevant perturbations~\cite{prb_daniele_2024}, such as a mixed valence limit, and $C_{3z}$ symmetry breaking corrections originating from strain or finite angular momentum in the hopping. These effects introduce a new length scale at low doping in addition to $k_F\sim \sqrt{x}^{-1}$ and establish a non-negligible Kondo temperature~\cite{prb_daniele_2024}. Additionally, the inclusion of holon fluctuations in the lower Hubbard band which enhances charge transfer processes leading to the paramagnetic Fermi liquid state. An in depth analysis of the competition between these terms is left to future studies. 

We conclude observing that the competing channels composed of $\bm\chi_0$ and $[\chi_{+\uparrow},\chi_{-\downarrow}]^{T}$ are coupled via a ferromagnetic exchange and their renormalization group flow follows the set of equations:
\begin{equation}
    \begin{split}
        &\frac{dJ_\perp}{d\log s}=J_\perp\frac{2\rho J^s_z + x m^*_{\rm W} J^p_z}{2},\\
        &\frac{dJ^s_z}{d\log s}=2\rho{J_\perp}^2,\quad \frac{dJ^p_z}{d\log s}= m^*_{\rm W}x {J_\perp}^2,
    \end{split}
\end{equation}
with bare coupling constants $J_\perp(0)=J_z/(2\sqrt{3})$, $J^s_z=-(J_z-\delta J_z)$, $J^p_z=-J_z/12$ and $\rho$ is the density of states of the itinerant band. For realistic values of the exchange constants the flow of theses couplings evolve towards the attractive fixed points forming a decoupled local moment. 

{\it Conclusions.---} In this paper we have shown that the interplay between Ising spin-orbit coupling and {\it conduction band} electron-electron repulsion in a doped magnetically ordered charge-transfer insulator gives rise to emergent canted, topological magnetic phases upon doping away from commensurate filling. Specifically, depending on the value of the Ising phase, we observe either interlayer ferromagnetic or antiferromagnetic correlations, which manifest as canted 120$^\circ$ ordering with anomalous Hall topological transport characteristics. We develop an analytical framework to elucidate the mechanisms driving these distinct magnetic correlations, supported by both small-cluster exact diagonalization and continuum model mean-field calculations. The continuum model also predicts the emergence of a quantum anomalous Hall canted $120^\circ$ phase through self-doping of the charge transfer band and the creation of holes in the lower Hubbard band at a filling factor $x=0$. Furthermore, our findings suggest the potential realization of spin polarons, which emerge as bound states between individual carriers in the itinerant band and local canting of the magnetic texture.
These spin polarons could be directly observable through probes sensitive to local charge distribution and out-of-plane spin polarization. 
Finally, we explore the competition between magnetism and the paramagnetic Fermi liquid state upon doping the itinerant layer in the single site limit. We predict two channels, one with $p$-wave structure leading to Kondo screening at a suppressed coupling scale. 

We conclude with a discussion of the insulating nature of the ferromagnetic phase and metallic nature of the heavy Fermi liquid that where observed experimentally. Incorporating potential disorder into each TMD monolayer we find that the ferromagnetic phase falls into the gaussian unitary ensemble  with broken time reversal symmetry, which Anderson localizes for infinitesimal disorder~\cite{RevModPhys.80.1355}. On the other hand, the $p$-wave nature of the heavy fermi liquid places the disordered model into the gaussian symplectic ensemble, which supports a metallic delocalized phase at weak to moderate disorder~\cite{RevModPhys.80.1355}. Both of these conclusions nicely capture the experimental transport properties.


{\it Acknowledgments.---} We acknowledge insightful discussions with Martin Claassen, Luca de Medici, Samuele Giuli, John Sous, Valentin Cr\'epel, Nick Bultinck, Antoine Georges, Juraj Hasik, Jason Kaye and Leon Balents, as well as Kin Fai mak and Jie Shan for stimulating our interest in these problems, describing unpublished data and helpful discussions. 
This work is partially supported by NSF Career Grant No.~DMR- 1941569 (J.H.P.). This work was  performed  in part at the Aspen Center for Physics, which is supported by National Science Foundation grant PHY-2210452 (J.H.P.) and as well as at the Kavli Institute of Theoretical Physics that is supported in part by the National Science Foundation under Grants No.~NSF PHY-1748958 and PHY-2309135 (D.G. and J.H.P.). 
The Flatiron Institute is a division of the Simons Foundation.

\bibliography{mainbib}

\begin{thebibliography}{44}%
\makeatletter
\providecommand \@ifxundefined [1]{%
 \@ifx{#1\undefined}
}%
\providecommand \@ifnum [1]{%
 \ifnum #1\expandafter \@firstoftwo
 \else \expandafter \@secondoftwo
 \fi
}%
\providecommand \@ifx [1]{%
 \ifx #1\expandafter \@firstoftwo
 \else \expandafter \@secondoftwo
 \fi
}%
\providecommand \natexlab [1]{#1}%
\providecommand \enquote  [1]{``#1''}%
\providecommand \bibnamefont  [1]{#1}%
\providecommand \bibfnamefont [1]{#1}%
\providecommand \citenamefont [1]{#1}%
\providecommand \href@noop [0]{\@secondoftwo}%
\providecommand \href [0]{\begingroup \@sanitize@url \@href}%
\providecommand \@href[1]{\@@startlink{#1}\@@href}%
\providecommand \@@href[1]{\endgroup#1\@@endlink}%
\providecommand \@sanitize@url [0]{\catcode `\\12\catcode `\$12\catcode
  `\&12\catcode `\#12\catcode `\^12\catcode `\_12\catcode `\%12\relax}%
\providecommand \@@startlink[1]{}%
\providecommand \@@endlink[0]{}%
\providecommand \url  [0]{\begingroup\@sanitize@url \@url }%
\providecommand \@url [1]{\endgroup\@href {#1}{\urlprefix }}%
\providecommand \urlprefix  [0]{URL }%
\providecommand \Eprint [0]{\href }%
\providecommand \doibase [0]{https://doi.org/}%
\providecommand \selectlanguage [0]{\@gobble}%
\providecommand \bibinfo  [0]{\@secondoftwo}%
\providecommand \bibfield  [0]{\@secondoftwo}%
\providecommand \translation [1]{[#1]}%
\providecommand \BibitemOpen [0]{}%
\providecommand \bibitemStop [0]{}%
\providecommand \bibitemNoStop [0]{.\EOS\space}%
\providecommand \EOS [0]{\spacefactor3000\relax}%
\providecommand \BibitemShut  [1]{\csname bibitem#1\endcsname}%
\let\auto@bib@innerbib\@empty
\bibitem [{\citenamefont {Andrei}\ \emph {et~al.}(2021)\citenamefont {Andrei},
  \citenamefont {Efetov}, \citenamefont {Jarillo-Herrero}, \citenamefont
  {MacDonald}, \citenamefont {Mak}, \citenamefont {Senthil}, \citenamefont
  {Tutuc}, \citenamefont {Yazdani},\ and\ \citenamefont
  {Young}}]{andrei2021marvels}%
  \BibitemOpen
  \bibfield  {author} {\bibinfo {author} {\bibfnamefont {E.~Y.}\ \bibnamefont
  {Andrei}}, \bibinfo {author} {\bibfnamefont {D.~K.}\ \bibnamefont {Efetov}},
  \bibinfo {author} {\bibfnamefont {P.}~\bibnamefont {Jarillo-Herrero}},
  \bibinfo {author} {\bibfnamefont {A.~H.}\ \bibnamefont {MacDonald}}, \bibinfo
  {author} {\bibfnamefont {K.~F.}\ \bibnamefont {Mak}}, \bibinfo {author}
  {\bibfnamefont {T.}~\bibnamefont {Senthil}}, \bibinfo {author} {\bibfnamefont
  {E.}~\bibnamefont {Tutuc}}, \bibinfo {author} {\bibfnamefont
  {A.}~\bibnamefont {Yazdani}},\ and\ \bibinfo {author} {\bibfnamefont {A.~F.}\
  \bibnamefont {Young}},\ }\bibfield  {title} {\bibinfo {title} {The marvels of
  moir{\'e} materials},\ }\href@noop {} {\bibfield  {journal} {\bibinfo
  {journal} {Nature Reviews Materials}\ }\textbf {\bibinfo {volume} {6}},\
  \bibinfo {pages} {201} (\bibinfo {year} {2021})}\BibitemShut {NoStop}%
\bibitem [{\citenamefont {Kennes}\ \emph {et~al.}(2021)\citenamefont {Kennes},
  \citenamefont {Claassen}, \citenamefont {Xian}, \citenamefont {Georges},
  \citenamefont {Millis}, \citenamefont {Hone}, \citenamefont {Dean},
  \citenamefont {Basov}, \citenamefont {Pasupathy},\ and\ \citenamefont
  {Rubio}}]{kennes2021moire}%
  \BibitemOpen
  \bibfield  {author} {\bibinfo {author} {\bibfnamefont {D.~M.}\ \bibnamefont
  {Kennes}}, \bibinfo {author} {\bibfnamefont {M.}~\bibnamefont {Claassen}},
  \bibinfo {author} {\bibfnamefont {L.}~\bibnamefont {Xian}}, \bibinfo {author}
  {\bibfnamefont {A.}~\bibnamefont {Georges}}, \bibinfo {author} {\bibfnamefont
  {A.~J.}\ \bibnamefont {Millis}}, \bibinfo {author} {\bibfnamefont
  {J.}~\bibnamefont {Hone}}, \bibinfo {author} {\bibfnamefont {C.~R.}\
  \bibnamefont {Dean}}, \bibinfo {author} {\bibfnamefont {D.}~\bibnamefont
  {Basov}}, \bibinfo {author} {\bibfnamefont {A.~N.}\ \bibnamefont
  {Pasupathy}},\ and\ \bibinfo {author} {\bibfnamefont {A.}~\bibnamefont
  {Rubio}},\ }\bibfield  {title} {\bibinfo {title} {Moir{\'e} heterostructures
  as a condensed-matter quantum simulator},\ }\href@noop {} {\bibfield
  {journal} {\bibinfo  {journal} {Nature Physics}\ }\textbf {\bibinfo {volume}
  {17}},\ \bibinfo {pages} {155} (\bibinfo {year} {2021})}\BibitemShut
  {NoStop}%
\bibitem [{\citenamefont {Mak}\ and\ \citenamefont
  {Shan}(2022)}]{Mak2022SemiconductorMM}%
  \BibitemOpen
  \bibfield  {author} {\bibinfo {author} {\bibfnamefont {K.~F.}\ \bibnamefont
  {Mak}}\ and\ \bibinfo {author} {\bibfnamefont {J.}~\bibnamefont {Shan}},\
  }\bibfield  {title} {\bibinfo {title} {Semiconductor moir{\'e} materials},\
  }\href@noop {} {\bibfield  {journal} {\bibinfo  {journal} {Nature
  Nanotechnology}\ }\textbf {\bibinfo {volume} {17}},\ \bibinfo {pages} {686 }
  (\bibinfo {year} {2022})}\BibitemShut {NoStop}%
\bibitem [{\citenamefont {Zhao}\ \emph {et~al.}(2023)\citenamefont {Zhao},
  \citenamefont {Shen}, \citenamefont {Tao}, \citenamefont {Han}, \citenamefont
  {Kang}, \citenamefont {Watanabe}, \citenamefont {Taniguchi}, \citenamefont
  {Mak},\ and\ \citenamefont {Shan}}]{Zhao_2023}%
  \BibitemOpen
  \bibfield  {author} {\bibinfo {author} {\bibfnamefont {W.}~\bibnamefont
  {Zhao}}, \bibinfo {author} {\bibfnamefont {B.}~\bibnamefont {Shen}}, \bibinfo
  {author} {\bibfnamefont {Z.}~\bibnamefont {Tao}}, \bibinfo {author}
  {\bibfnamefont {Z.}~\bibnamefont {Han}}, \bibinfo {author} {\bibfnamefont
  {K.}~\bibnamefont {Kang}}, \bibinfo {author} {\bibfnamefont {K.}~\bibnamefont
  {Watanabe}}, \bibinfo {author} {\bibfnamefont {T.}~\bibnamefont {Taniguchi}},
  \bibinfo {author} {\bibfnamefont {K.~F.}\ \bibnamefont {Mak}},\ and\ \bibinfo
  {author} {\bibfnamefont {J.}~\bibnamefont {Shan}},\ }\bibfield  {title}
  {\bibinfo {title} {Gate-tunable heavy fermions in a moiré kondo lattice},\
  }\href {https://doi.org/10.1038/s41586-023-05800-7} {\bibfield  {journal}
  {\bibinfo  {journal} {Nature}\ }\textbf {\bibinfo {volume} {616}},\ \bibinfo
  {pages} {61–65} (\bibinfo {year} {2023})}\BibitemShut {NoStop}%
\bibitem [{\citenamefont {Zhao}\ \emph {et~al.}(2024)\citenamefont {Zhao},
  \citenamefont {Shen}, \citenamefont {Tao}, \citenamefont {Kim}, \citenamefont
  {Kn{\"u}ppel}, \citenamefont {Han}, \citenamefont {Zhang}, \citenamefont
  {Watanabe}, \citenamefont {Taniguchi}, \citenamefont {Chowdhury} \emph
  {et~al.}}]{zhao2024emergence}%
  \BibitemOpen
  \bibfield  {author} {\bibinfo {author} {\bibfnamefont {W.}~\bibnamefont
  {Zhao}}, \bibinfo {author} {\bibfnamefont {B.}~\bibnamefont {Shen}}, \bibinfo
  {author} {\bibfnamefont {Z.}~\bibnamefont {Tao}}, \bibinfo {author}
  {\bibfnamefont {S.}~\bibnamefont {Kim}}, \bibinfo {author} {\bibfnamefont
  {P.}~\bibnamefont {Kn{\"u}ppel}}, \bibinfo {author} {\bibfnamefont
  {Z.}~\bibnamefont {Han}}, \bibinfo {author} {\bibfnamefont {Y.}~\bibnamefont
  {Zhang}}, \bibinfo {author} {\bibfnamefont {K.}~\bibnamefont {Watanabe}},
  \bibinfo {author} {\bibfnamefont {T.}~\bibnamefont {Taniguchi}}, \bibinfo
  {author} {\bibfnamefont {D.}~\bibnamefont {Chowdhury}}, \emph {et~al.},\
  }\bibfield  {title} {\bibinfo {title} {Emergence of ferromagnetism at the
  onset of moir{\'e} kondo breakdown},\ }\href@noop {} {\bibfield  {journal}
  {\bibinfo  {journal} {Nature Physics}\ ,\ \bibinfo {pages} {1}} (\bibinfo
  {year} {2024})}\BibitemShut {NoStop}%
\bibitem [{\citenamefont {Guerci}\ \emph {et~al.}(2023)\citenamefont {Guerci},
  \citenamefont {Wang}, \citenamefont {Zang}, \citenamefont {Cano},
  \citenamefont {Pixley},\ and\ \citenamefont {Millis}}]{Guerci_2023}%
  \BibitemOpen
  \bibfield  {author} {\bibinfo {author} {\bibfnamefont {D.}~\bibnamefont
  {Guerci}}, \bibinfo {author} {\bibfnamefont {J.}~\bibnamefont {Wang}},
  \bibinfo {author} {\bibfnamefont {J.}~\bibnamefont {Zang}}, \bibinfo {author}
  {\bibfnamefont {J.}~\bibnamefont {Cano}}, \bibinfo {author} {\bibfnamefont
  {J.~H.}\ \bibnamefont {Pixley}},\ and\ \bibinfo {author} {\bibfnamefont
  {A.}~\bibnamefont {Millis}},\ }\bibfield  {title} {\bibinfo {title} {Chiral
  kondo lattice in doped mote2/wse2 bilayers},\ }\bibfield  {journal} {\bibinfo
   {journal} {Science Advances}\ }\textbf {\bibinfo {volume} {9}},\ \href
  {https://doi.org/10.1126/sciadv.ade7701} {10.1126/sciadv.ade7701} (\bibinfo
  {year} {2023})\BibitemShut {NoStop}%
\bibitem [{\citenamefont {Guerci}\ \emph {et~al.}(2024)\citenamefont {Guerci},
  \citenamefont {Lucht}, \citenamefont {Cr\'epel}, \citenamefont {Cano},
  \citenamefont {Pixley},\ and\ \citenamefont {Millis}}]{prb_daniele_2024}%
  \BibitemOpen
  \bibfield  {author} {\bibinfo {author} {\bibfnamefont {D.}~\bibnamefont
  {Guerci}}, \bibinfo {author} {\bibfnamefont {K.~P.}\ \bibnamefont {Lucht}},
  \bibinfo {author} {\bibfnamefont {V.}~\bibnamefont {Cr\'epel}}, \bibinfo
  {author} {\bibfnamefont {J.}~\bibnamefont {Cano}}, \bibinfo {author}
  {\bibfnamefont {J.~H.}\ \bibnamefont {Pixley}},\ and\ \bibinfo {author}
  {\bibfnamefont {A.}~\bibnamefont {Millis}},\ }\bibfield  {title} {\bibinfo
  {title} {Topological kondo semimetal and insulator in ab-stacked
  heterobilayer transition metal dichalcogenides},\ }\href
  {https://doi.org/10.1103/PhysRevB.110.165128} {\bibfield  {journal} {\bibinfo
   {journal} {Phys. Rev. B}\ }\textbf {\bibinfo {volume} {110}},\ \bibinfo
  {pages} {165128} (\bibinfo {year} {2024})}\BibitemShut {NoStop}%
\bibitem [{\citenamefont {Dalal}\ and\ \citenamefont
  {Ruhman}(2021)}]{Dalal_2021}%
  \BibitemOpen
  \bibfield  {author} {\bibinfo {author} {\bibfnamefont {A.}~\bibnamefont
  {Dalal}}\ and\ \bibinfo {author} {\bibfnamefont {J.}~\bibnamefont {Ruhman}},\
  }\bibfield  {title} {\bibinfo {title} {Orbitally selective mott phase in
  electron-doped twisted transition metal-dichalcogenides: A possible
  realization of the kondo lattice model},\ }\href
  {https://doi.org/10.1103/PhysRevResearch.3.043173} {\bibfield  {journal}
  {\bibinfo  {journal} {Phys. Rev. Research}\ }\textbf {\bibinfo {volume}
  {3}},\ \bibinfo {pages} {043173} (\bibinfo {year} {2021})}\BibitemShut
  {NoStop}%
\bibitem [{\citenamefont {Kumar}\ \emph {et~al.}(2022)\citenamefont {Kumar},
  \citenamefont {Hu}, \citenamefont {MacDonald},\ and\ \citenamefont
  {Potter}}]{Kumar_2022}%
  \BibitemOpen
  \bibfield  {author} {\bibinfo {author} {\bibfnamefont {A.}~\bibnamefont
  {Kumar}}, \bibinfo {author} {\bibfnamefont {N.~C.}\ \bibnamefont {Hu}},
  \bibinfo {author} {\bibfnamefont {A.~H.}\ \bibnamefont {MacDonald}},\ and\
  \bibinfo {author} {\bibfnamefont {A.~C.}\ \bibnamefont {Potter}},\ }\bibfield
   {title} {\bibinfo {title} {Gate-tunable heavy fermion quantum criticality in
  a moir\'e kondo lattice},\ }\href
  {https://doi.org/10.1103/PhysRevB.106.L041116} {\bibfield  {journal}
  {\bibinfo  {journal} {Phys. Rev. B}\ }\textbf {\bibinfo {volume} {106}},\
  \bibinfo {pages} {L041116} (\bibinfo {year} {2022})}\BibitemShut {NoStop}%
\bibitem [{\citenamefont {Xie}\ \emph {et~al.}(2024)\citenamefont {Xie},
  \citenamefont {Chen}, \citenamefont {Fang},\ and\ \citenamefont
  {Si}}]{xie2024topologicalkondosemimetalsemulated}%
  \BibitemOpen
  \bibfield  {author} {\bibinfo {author} {\bibfnamefont {F.}~\bibnamefont
  {Xie}}, \bibinfo {author} {\bibfnamefont {L.}~\bibnamefont {Chen}}, \bibinfo
  {author} {\bibfnamefont {Y.}~\bibnamefont {Fang}},\ and\ \bibinfo {author}
  {\bibfnamefont {Q.}~\bibnamefont {Si}},\ }\href
  {https://arxiv.org/abs/2407.20182} {\bibinfo {title} {Topological kondo
  semimetals emulated in hetero-bilayer transition metal dichalcogenides}}
  (\bibinfo {year} {2024}),\ \Eprint {https://arxiv.org/abs/2407.20182}
  {arXiv:2407.20182 [cond-mat.str-el]} \BibitemShut {NoStop}%
\bibitem [{\citenamefont {Mendez-Valderrama}\ \emph {et~al.}(2024)\citenamefont
  {Mendez-Valderrama}, \citenamefont {Kim},\ and\ \citenamefont
  {Chowdhury}}]{mendezvalderrama2024correlatedtopologicalmixedvalenceinsulators}%
  \BibitemOpen
  \bibfield  {author} {\bibinfo {author} {\bibfnamefont {J.~F.}\ \bibnamefont
  {Mendez-Valderrama}}, \bibinfo {author} {\bibfnamefont {S.}~\bibnamefont
  {Kim}},\ and\ \bibinfo {author} {\bibfnamefont {D.}~\bibnamefont
  {Chowdhury}},\ }\href {https://arxiv.org/abs/2407.14583} {\bibinfo {title}
  {Correlated topological mixed-valence insulators in moir\'e hetero-bilayers}}
  (\bibinfo {year} {2024}),\ \Eprint {https://arxiv.org/abs/2407.14583}
  {arXiv:2407.14583 [cond-mat.str-el]} \BibitemShut {NoStop}%
\bibitem [{\citenamefont {Coleman}\ \emph {et~al.}(2001)\citenamefont
  {Coleman}, \citenamefont {Pépin}, \citenamefont {Si},\ and\ \citenamefont
  {Ramazashvili}}]{Coleman_2001}%
  \BibitemOpen
  \bibfield  {author} {\bibinfo {author} {\bibfnamefont {P.}~\bibnamefont
  {Coleman}}, \bibinfo {author} {\bibfnamefont {C.}~\bibnamefont {Pépin}},
  \bibinfo {author} {\bibfnamefont {Q.}~\bibnamefont {Si}},\ and\ \bibinfo
  {author} {\bibfnamefont {R.}~\bibnamefont {Ramazashvili}},\ }\bibfield
  {title} {\bibinfo {title} {How do fermi liquids get heavy and die?},\ }\href
  {https://doi.org/10.1088/0953-8984/13/35/202} {\bibfield  {journal} {\bibinfo
   {journal} {Journal of Physics: Condensed Matter}\ }\textbf {\bibinfo
  {volume} {13}},\ \bibinfo {pages} {R723–R738} (\bibinfo {year}
  {2001})}\BibitemShut {NoStop}%
\bibitem [{\citenamefont {Senthil}\ \emph {et~al.}(2004)\citenamefont
  {Senthil}, \citenamefont {Vojta},\ and\ \citenamefont
  {Sachdev}}]{Senthil_2004}%
  \BibitemOpen
  \bibfield  {author} {\bibinfo {author} {\bibfnamefont {T.}~\bibnamefont
  {Senthil}}, \bibinfo {author} {\bibfnamefont {M.}~\bibnamefont {Vojta}},\
  and\ \bibinfo {author} {\bibfnamefont {S.}~\bibnamefont {Sachdev}},\
  }\bibfield  {title} {\bibinfo {title} {Weak magnetism and non-fermi liquids
  near heavy-fermion critical points},\ }\href
  {https://doi.org/10.1103/PhysRevB.69.035111} {\bibfield  {journal} {\bibinfo
  {journal} {Phys. Rev. B}\ }\textbf {\bibinfo {volume} {69}},\ \bibinfo
  {pages} {035111} (\bibinfo {year} {2004})}\BibitemShut {NoStop}%
\bibitem [{\citenamefont {Biermann}\ \emph {et~al.}(2005)\citenamefont
  {Biermann}, \citenamefont {de' Medici},\ and\ \citenamefont
  {Georges}}]{Biermann_2005}%
  \BibitemOpen
  \bibfield  {author} {\bibinfo {author} {\bibfnamefont {S.}~\bibnamefont
  {Biermann}}, \bibinfo {author} {\bibfnamefont {L.}~\bibnamefont {de'
  Medici}},\ and\ \bibinfo {author} {\bibfnamefont {A.}~\bibnamefont
  {Georges}},\ }\bibfield  {title} {\bibinfo {title} {Non-fermi-liquid behavior
  and double-exchange physics in orbital-selective mott systems},\ }\href
  {https://doi.org/10.1103/PhysRevLett.95.206401} {\bibfield  {journal}
  {\bibinfo  {journal} {Phys. Rev. Lett.}\ }\textbf {\bibinfo {volume} {95}},\
  \bibinfo {pages} {206401} (\bibinfo {year} {2005})}\BibitemShut {NoStop}%
\bibitem [{\citenamefont {Yamamoto}\ and\ \citenamefont
  {Si}(2007)}]{Yamamoto_2007}%
  \BibitemOpen
  \bibfield  {author} {\bibinfo {author} {\bibfnamefont {S.~J.}\ \bibnamefont
  {Yamamoto}}\ and\ \bibinfo {author} {\bibfnamefont {Q.}~\bibnamefont {Si}},\
  }\bibfield  {title} {\bibinfo {title} {Fermi surface and antiferromagnetism
  in the kondo lattice: An asymptotically exact solution in $d>1$ dimensions},\
  }\href {https://doi.org/10.1103/PhysRevLett.99.016401} {\bibfield  {journal}
  {\bibinfo  {journal} {Phys. Rev. Lett.}\ }\textbf {\bibinfo {volume} {99}},\
  \bibinfo {pages} {016401} (\bibinfo {year} {2007})}\BibitemShut {NoStop}%
\bibitem [{\citenamefont {Koga}\ \emph {et~al.}(2008)\citenamefont {Koga},
  \citenamefont {Kawakami}, \citenamefont {Peters},\ and\ \citenamefont
  {Pruschke}}]{Koga_2008}%
  \BibitemOpen
  \bibfield  {author} {\bibinfo {author} {\bibfnamefont {A.}~\bibnamefont
  {Koga}}, \bibinfo {author} {\bibfnamefont {N.}~\bibnamefont {Kawakami}},
  \bibinfo {author} {\bibfnamefont {R.}~\bibnamefont {Peters}},\ and\ \bibinfo
  {author} {\bibfnamefont {T.}~\bibnamefont {Pruschke}},\ }\bibfield  {title}
  {\bibinfo {title} {Magnetic properties of the extended periodic anderson
  model},\ }\href {https://doi.org/10.1143/jpsj.77.033704} {\bibfield
  {journal} {\bibinfo  {journal} {Journal of the Physical Society of Japan}\
  }\textbf {\bibinfo {volume} {77}},\ \bibinfo {pages} {033704} (\bibinfo
  {year} {2008})}\BibitemShut {NoStop}%
\bibitem [{\citenamefont {Li}\ \emph {et~al.}(2021)\citenamefont {Li},
  \citenamefont {Jiang}, \citenamefont {Shen}, \citenamefont {Zhang},
  \citenamefont {Li}, \citenamefont {Tao}, \citenamefont {Devakul},
  \citenamefont {Watanabe}, \citenamefont {Taniguchi}, \citenamefont {Fu},
  \citenamefont {Shan},\ and\ \citenamefont {Mak}}]{Li_2021}%
  \BibitemOpen
  \bibfield  {author} {\bibinfo {author} {\bibfnamefont {T.}~\bibnamefont
  {Li}}, \bibinfo {author} {\bibfnamefont {S.}~\bibnamefont {Jiang}}, \bibinfo
  {author} {\bibfnamefont {B.}~\bibnamefont {Shen}}, \bibinfo {author}
  {\bibfnamefont {Y.}~\bibnamefont {Zhang}}, \bibinfo {author} {\bibfnamefont
  {L.}~\bibnamefont {Li}}, \bibinfo {author} {\bibfnamefont {Z.}~\bibnamefont
  {Tao}}, \bibinfo {author} {\bibfnamefont {T.}~\bibnamefont {Devakul}},
  \bibinfo {author} {\bibfnamefont {K.}~\bibnamefont {Watanabe}}, \bibinfo
  {author} {\bibfnamefont {T.}~\bibnamefont {Taniguchi}}, \bibinfo {author}
  {\bibfnamefont {L.}~\bibnamefont {Fu}}, \bibinfo {author} {\bibfnamefont
  {J.}~\bibnamefont {Shan}},\ and\ \bibinfo {author} {\bibfnamefont {K.~F.}\
  \bibnamefont {Mak}},\ }\bibfield  {title} {\bibinfo {title} {Quantum
  anomalous hall effect from intertwined moir{\'{e}} bands},\ }\href
  {https://doi.org/10.1038/s41586-021-04171-1} {\bibfield  {journal} {\bibinfo
  {journal} {Nature}\ }\textbf {\bibinfo {volume} {600}},\ \bibinfo {pages}
  {641} (\bibinfo {year} {2021})}\BibitemShut {NoStop}%
\bibitem [{\citenamefont {Jia}\ \emph {et~al.}(2023)\citenamefont {Jia},
  \citenamefont {Ma}, \citenamefont {Luo},\ and\ \citenamefont
  {Chen}}]{PhysRevResearch.5.L042033}%
  \BibitemOpen
  \bibfield  {author} {\bibinfo {author} {\bibfnamefont {H.}~\bibnamefont
  {Jia}}, \bibinfo {author} {\bibfnamefont {B.}~\bibnamefont {Ma}}, \bibinfo
  {author} {\bibfnamefont {R.~L.}\ \bibnamefont {Luo}},\ and\ \bibinfo {author}
  {\bibfnamefont {G.}~\bibnamefont {Chen}},\ }\bibfield  {title} {\bibinfo
  {title} {Double exchange, itinerant ferromagnetism, and topological hall
  effect in moir\'e heterobilayer},\ }\href
  {https://doi.org/10.1103/PhysRevResearch.5.L042033} {\bibfield  {journal}
  {\bibinfo  {journal} {Phys. Rev. Res.}\ }\textbf {\bibinfo {volume} {5}},\
  \bibinfo {pages} {L042033} (\bibinfo {year} {2023})}\BibitemShut {NoStop}%
\bibitem [{\citenamefont {Seifert}\ and\ \citenamefont
  {Balents}(2024)}]{Seifert}%
  \BibitemOpen
  \bibfield  {author} {\bibinfo {author} {\bibfnamefont {U.~F.~P.}\
  \bibnamefont {Seifert}}\ and\ \bibinfo {author} {\bibfnamefont
  {L.}~\bibnamefont {Balents}},\ }\bibfield  {title} {\bibinfo {title} {Spin
  polarons and ferromagnetism in doped dilute moir\'e-mott insulators},\ }\href
  {https://doi.org/10.1103/PhysRevLett.132.046501} {\bibfield  {journal}
  {\bibinfo  {journal} {Phys. Rev. Lett.}\ }\textbf {\bibinfo {volume} {132}},\
  \bibinfo {pages} {046501} (\bibinfo {year} {2024})}\BibitemShut {NoStop}%
\bibitem [{\citenamefont {Korm{\'a}nyos}\ \emph {et~al.}(2015)\citenamefont
  {Korm{\'a}nyos}, \citenamefont {Burkard}, \citenamefont {Gmitra},
  \citenamefont {Fabian}, \citenamefont {Z{\'o}lyomi}, \citenamefont
  {Drummond},\ and\ \citenamefont {Fal’ko}}]{kormanyos2015k}%
  \BibitemOpen
  \bibfield  {author} {\bibinfo {author} {\bibfnamefont {A.}~\bibnamefont
  {Korm{\'a}nyos}}, \bibinfo {author} {\bibfnamefont {G.}~\bibnamefont
  {Burkard}}, \bibinfo {author} {\bibfnamefont {M.}~\bibnamefont {Gmitra}},
  \bibinfo {author} {\bibfnamefont {J.}~\bibnamefont {Fabian}}, \bibinfo
  {author} {\bibfnamefont {V.}~\bibnamefont {Z{\'o}lyomi}}, \bibinfo {author}
  {\bibfnamefont {N.~D.}\ \bibnamefont {Drummond}},\ and\ \bibinfo {author}
  {\bibfnamefont {V.}~\bibnamefont {Fal’ko}},\ }\bibfield  {title} {\bibinfo
  {title} {k{\textperiodcentered} p theory for two-dimensional transition metal
  dichalcogenide semiconductors},\ }\href@noop {} {\bibfield  {journal}
  {\bibinfo  {journal} {2D Materials}\ }\textbf {\bibinfo {volume} {2}},\
  \bibinfo {pages} {022001} (\bibinfo {year} {2015})}\BibitemShut {NoStop}%
\bibitem [{\citenamefont {Zhang}\ \emph {et~al.}(2021)\citenamefont {Zhang},
  \citenamefont {Devakul},\ and\ \citenamefont {Fu}}]{Zhang_2021}%
  \BibitemOpen
  \bibfield  {author} {\bibinfo {author} {\bibfnamefont {Y.}~\bibnamefont
  {Zhang}}, \bibinfo {author} {\bibfnamefont {T.}~\bibnamefont {Devakul}},\
  and\ \bibinfo {author} {\bibfnamefont {L.}~\bibnamefont {Fu}},\ }\bibfield
  {title} {\bibinfo {title} {Spin-textured chern bands in {AB}-stacked
  transition metal dichalcogenide bilayers},\ }\bibfield  {journal} {\bibinfo
  {journal} {Proceedings of the National Academy of Sciences}\ }\textbf
  {\bibinfo {volume} {118}},\ \href {https://doi.org/10.1073/pnas.2112673118}
  {10.1073/pnas.2112673118} (\bibinfo {year} {2021})\BibitemShut {NoStop}%
\bibitem [{\citenamefont {Tao}\ \emph {et~al.}(2024)\citenamefont {Tao},
  \citenamefont {Shen}, \citenamefont {Jiang}, \citenamefont {Li},
  \citenamefont {Li}, \citenamefont {Ma}, \citenamefont {Zhao}, \citenamefont
  {Hu}, \citenamefont {Pistunova}, \citenamefont {Watanabe}, \citenamefont
  {Taniguchi}, \citenamefont {Heinz}, \citenamefont {Mak},\ and\ \citenamefont
  {Shan}}]{tao2022valleycoherent}%
  \BibitemOpen
  \bibfield  {author} {\bibinfo {author} {\bibfnamefont {Z.}~\bibnamefont
  {Tao}}, \bibinfo {author} {\bibfnamefont {B.}~\bibnamefont {Shen}}, \bibinfo
  {author} {\bibfnamefont {S.}~\bibnamefont {Jiang}}, \bibinfo {author}
  {\bibfnamefont {T.}~\bibnamefont {Li}}, \bibinfo {author} {\bibfnamefont
  {L.}~\bibnamefont {Li}}, \bibinfo {author} {\bibfnamefont {L.}~\bibnamefont
  {Ma}}, \bibinfo {author} {\bibfnamefont {W.}~\bibnamefont {Zhao}}, \bibinfo
  {author} {\bibfnamefont {J.}~\bibnamefont {Hu}}, \bibinfo {author}
  {\bibfnamefont {K.}~\bibnamefont {Pistunova}}, \bibinfo {author}
  {\bibfnamefont {K.}~\bibnamefont {Watanabe}}, \bibinfo {author}
  {\bibfnamefont {T.}~\bibnamefont {Taniguchi}}, \bibinfo {author}
  {\bibfnamefont {T.~F.}\ \bibnamefont {Heinz}}, \bibinfo {author}
  {\bibfnamefont {K.~F.}\ \bibnamefont {Mak}},\ and\ \bibinfo {author}
  {\bibfnamefont {J.}~\bibnamefont {Shan}},\ }\bibfield  {title} {\bibinfo
  {title} {Valley-coherent quantum anomalous hall state in ab-stacked
  ${\mathrm{mote}}_{2}/{\mathrm{w}\mathrm{s}\mathrm{e}}_{2}$ bilayers},\ }\href
  {https://doi.org/10.1103/PhysRevX.14.011004} {\bibfield  {journal} {\bibinfo
  {journal} {Phys. Rev. X}\ }\textbf {\bibinfo {volume} {14}},\ \bibinfo
  {pages} {011004} (\bibinfo {year} {2024})}\BibitemShut {NoStop}%
\bibitem [{\citenamefont {Devakul}\ and\ \citenamefont
  {Fu}(2022)}]{Devakul_2021}%
  \BibitemOpen
  \bibfield  {author} {\bibinfo {author} {\bibfnamefont {T.}~\bibnamefont
  {Devakul}}\ and\ \bibinfo {author} {\bibfnamefont {L.}~\bibnamefont {Fu}},\
  }\bibfield  {title} {\bibinfo {title} {Quantum anomalous hall effect from
  inverted charge transfer gap},\ }\bibfield  {journal} {\bibinfo  {journal}
  {Physical Review X}\ }\textbf {\bibinfo {volume} {12}},\ \href
  {https://doi.org/10.1103/physrevx.12.021031} {10.1103/physrevx.12.021031}
  (\bibinfo {year} {2022})\BibitemShut {NoStop}%
\bibitem [{\citenamefont {Xie}\ \emph {et~al.}(2022)\citenamefont {Xie},
  \citenamefont {Zhang}, \citenamefont {Hu}, \citenamefont {Mak},\ and\
  \citenamefont {Law}}]{KTLaw_2022}%
  \BibitemOpen
  \bibfield  {author} {\bibinfo {author} {\bibfnamefont {Y.-M.}\ \bibnamefont
  {Xie}}, \bibinfo {author} {\bibfnamefont {C.-P.}\ \bibnamefont {Zhang}},
  \bibinfo {author} {\bibfnamefont {J.-X.}\ \bibnamefont {Hu}}, \bibinfo
  {author} {\bibfnamefont {K.~F.}\ \bibnamefont {Mak}},\ and\ \bibinfo {author}
  {\bibfnamefont {K.~T.}\ \bibnamefont {Law}},\ }\bibfield  {title} {\bibinfo
  {title} {Valley-polarized quantum anomalous hall state in moir\'e
  ${\mathrm{mote}}_{2}/{\mathrm{wse}}_{2}$ heterobilayers},\ }\href
  {https://doi.org/10.1103/PhysRevLett.128.026402} {\bibfield  {journal}
  {\bibinfo  {journal} {Phys. Rev. Lett.}\ }\textbf {\bibinfo {volume} {128}},\
  \bibinfo {pages} {026402} (\bibinfo {year} {2022})}\BibitemShut {NoStop}%
\bibitem [{\citenamefont {Dong}\ and\ \citenamefont {Zhang}(2023)}]{Dong_2023}%
  \BibitemOpen
  \bibfield  {author} {\bibinfo {author} {\bibfnamefont {Z.}~\bibnamefont
  {Dong}}\ and\ \bibinfo {author} {\bibfnamefont {Y.-H.}\ \bibnamefont
  {Zhang}},\ }\bibfield  {title} {\bibinfo {title} {Excitonic chern insulator
  and kinetic ferromagnetism in a ${\mathrm{mote}}_{2}/{\mathrm{wse}}_{2}$
  moir\'e bilayer},\ }\href {https://doi.org/10.1103/PhysRevB.107.L081101}
  {\bibfield  {journal} {\bibinfo  {journal} {Phys. Rev. B}\ }\textbf {\bibinfo
  {volume} {107}},\ \bibinfo {pages} {L081101} (\bibinfo {year}
  {2023})}\BibitemShut {NoStop}%
\bibitem [{\citenamefont {Anderson}(1970)}]{Anderson_1970}%
  \BibitemOpen
  \bibfield  {author} {\bibinfo {author} {\bibfnamefont {P.}~\bibnamefont
  {Anderson}},\ }\bibfield  {title} {\bibinfo {title} {A poor man's derivation
  of scaling laws for the kondo problem},\ }\href
  {https://doi.org/10.1088/0022-3719/3/12/008} {\bibfield  {journal} {\bibinfo
  {journal} {Journal of Physics C: Solid State Physics}\ }\textbf {\bibinfo
  {volume} {3}},\ \bibinfo {pages} {2436} (\bibinfo {year} {1970})}\BibitemShut
  {NoStop}%
\bibitem [{sup()}]{supplementary}%
  \BibitemOpen
  \href@noop {} {}\bibinfo {note} {See Supplementary Material at url ... for
  details on the continuum and the tight binding Hamiltonian, magnetic ground
  state and fluctuations of the N\'eel ordered state.}\BibitemShut {Stop}%
\bibitem [{\citenamefont {Haldane}(2004)}]{Haldane_2004}%
  \BibitemOpen
  \bibfield  {author} {\bibinfo {author} {\bibfnamefont {F.~D.~M.}\
  \bibnamefont {Haldane}},\ }\bibfield  {title} {\bibinfo {title} {Berry
  curvature on the fermi surface: Anomalous hall effect as a topological
  fermi-liquid property},\ }\bibfield  {journal} {\bibinfo  {journal} {Physical
  Review Letters}\ }\textbf {\bibinfo {volume} {93}},\ \href
  {https://doi.org/10.1103/physrevlett.93.206602}
  {10.1103/physrevlett.93.206602} (\bibinfo {year} {2004})\BibitemShut
  {NoStop}%
\bibitem [{\citenamefont {Binz}\ and\ \citenamefont
  {Vishwanath}(2008)}]{Binz_2008}%
  \BibitemOpen
  \bibfield  {author} {\bibinfo {author} {\bibfnamefont {B.}~\bibnamefont
  {Binz}}\ and\ \bibinfo {author} {\bibfnamefont {A.}~\bibnamefont
  {Vishwanath}},\ }\bibfield  {title} {\bibinfo {title} {Chirality induced
  anomalous-hall effect in helical spin crystals},\ }\href
  {https://doi.org/10.1016/j.physb.2007.10.136} {\bibfield  {journal} {\bibinfo
   {journal} {Physica B: Condensed Matter}\ }\textbf {\bibinfo {volume}
  {403}},\ \bibinfo {pages} {1336–1340} (\bibinfo {year} {2008})}\BibitemShut
  {NoStop}%
\bibitem [{\citenamefont {Verma}\ \emph {et~al.}(2022)\citenamefont {Verma},
  \citenamefont {Addison},\ and\ \citenamefont {Randeria}}]{verma2022unified}%
  \BibitemOpen
  \bibfield  {author} {\bibinfo {author} {\bibfnamefont {N.}~\bibnamefont
  {Verma}}, \bibinfo {author} {\bibfnamefont {Z.}~\bibnamefont {Addison}},\
  and\ \bibinfo {author} {\bibfnamefont {M.}~\bibnamefont {Randeria}},\
  }\href@noop {} {\bibinfo {title} {Unified theory of the anomalous and
  topological hall effects with phase space berry curvatures}} (\bibinfo {year}
  {2022}),\ \Eprint {https://arxiv.org/abs/2203.07356} {arXiv:2203.07356
  [cond-mat.mes-hall]} \BibitemShut {NoStop}%
\bibitem [{\citenamefont {Martin}\ and\ \citenamefont
  {Batista}(2008)}]{IvarMartin_20108}%
  \BibitemOpen
  \bibfield  {author} {\bibinfo {author} {\bibfnamefont {I.}~\bibnamefont
  {Martin}}\ and\ \bibinfo {author} {\bibfnamefont {C.~D.}\ \bibnamefont
  {Batista}},\ }\bibfield  {title} {\bibinfo {title} {Itinerant electron-driven
  chiral magnetic ordering and spontaneous quantum hall effect in triangular
  lattice models},\ }\href {https://doi.org/10.1103/PhysRevLett.101.156402}
  {\bibfield  {journal} {\bibinfo  {journal} {Phys. Rev. Lett.}\ }\textbf
  {\bibinfo {volume} {101}},\ \bibinfo {pages} {156402} (\bibinfo {year}
  {2008})}\BibitemShut {NoStop}%
\bibitem [{\citenamefont {Shraiman}\ and\ \citenamefont
  {Siggia}(1988{\natexlab{a}})}]{PhysRevLett.61.467}%
  \BibitemOpen
  \bibfield  {author} {\bibinfo {author} {\bibfnamefont {B.~I.}\ \bibnamefont
  {Shraiman}}\ and\ \bibinfo {author} {\bibfnamefont {E.~D.}\ \bibnamefont
  {Siggia}},\ }\bibfield  {title} {\bibinfo {title} {Mobile vacancies in a
  quantum heisenberg antiferromagnet},\ }\href
  {https://doi.org/10.1103/PhysRevLett.61.467} {\bibfield  {journal} {\bibinfo
  {journal} {Phys. Rev. Lett.}\ }\textbf {\bibinfo {volume} {61}},\ \bibinfo
  {pages} {467} (\bibinfo {year} {1988}{\natexlab{a}})}\BibitemShut {NoStop}%
\bibitem [{\citenamefont {Shastry}\ and\ \citenamefont
  {Mattis}(1981)}]{Shastry_1981}%
  \BibitemOpen
  \bibfield  {author} {\bibinfo {author} {\bibfnamefont {B.~S.}\ \bibnamefont
  {Shastry}}\ and\ \bibinfo {author} {\bibfnamefont {D.~C.}\ \bibnamefont
  {Mattis}},\ }\bibfield  {title} {\bibinfo {title} {Theory of the magnetic
  polaron},\ }\href {https://doi.org/10.1103/PhysRevB.24.5340} {\bibfield
  {journal} {\bibinfo  {journal} {Phys. Rev. B}\ }\textbf {\bibinfo {volume}
  {24}},\ \bibinfo {pages} {5340} (\bibinfo {year} {1981})}\BibitemShut
  {NoStop}%
\bibitem [{\citenamefont {Shraiman}\ and\ \citenamefont
  {Siggia}(1988{\natexlab{b}})}]{Siggia_1988}%
  \BibitemOpen
  \bibfield  {author} {\bibinfo {author} {\bibfnamefont {B.~I.}\ \bibnamefont
  {Shraiman}}\ and\ \bibinfo {author} {\bibfnamefont {E.~D.}\ \bibnamefont
  {Siggia}},\ }\bibfield  {title} {\bibinfo {title} {Two-particle excitations
  in antiferromagnetic insulators},\ }\href
  {https://doi.org/10.1103/PhysRevLett.60.740} {\bibfield  {journal} {\bibinfo
  {journal} {Phys. Rev. Lett.}\ }\textbf {\bibinfo {volume} {60}},\ \bibinfo
  {pages} {740} (\bibinfo {year} {1988}{\natexlab{b}})}\BibitemShut {NoStop}%
\bibitem [{\citenamefont {Sigrist}\ \emph {et~al.}(1991)\citenamefont
  {Sigrist}, \citenamefont {Tsunetsuga},\ and\ \citenamefont
  {Ueda}}]{Sigrist_1991}%
  \BibitemOpen
  \bibfield  {author} {\bibinfo {author} {\bibfnamefont {M.}~\bibnamefont
  {Sigrist}}, \bibinfo {author} {\bibfnamefont {H.}~\bibnamefont
  {Tsunetsuga}},\ and\ \bibinfo {author} {\bibfnamefont {K.}~\bibnamefont
  {Ueda}},\ }\bibfield  {title} {\bibinfo {title} {Rigorous results for the
  one-electron kondo-lattice model},\ }\href
  {https://doi.org/10.1103/PhysRevLett.67.2211} {\bibfield  {journal} {\bibinfo
   {journal} {Phys. Rev. Lett.}\ }\textbf {\bibinfo {volume} {67}},\ \bibinfo
  {pages} {2211} (\bibinfo {year} {1991})}\BibitemShut {NoStop}%
\bibitem [{\citenamefont {Dombre}\ and\ \citenamefont {Read}(1989)}]{read1989}%
  \BibitemOpen
  \bibfield  {author} {\bibinfo {author} {\bibfnamefont {T.}~\bibnamefont
  {Dombre}}\ and\ \bibinfo {author} {\bibfnamefont {N.}~\bibnamefont {Read}},\
  }\bibfield  {title} {\bibinfo {title} {Nonlinear \ensuremath{\sigma} models
  for triangular quantum antiferromagnets},\ }\href
  {https://doi.org/10.1103/PhysRevB.39.6797} {\bibfield  {journal} {\bibinfo
  {journal} {Phys. Rev. B}\ }\textbf {\bibinfo {volume} {39}},\ \bibinfo
  {pages} {6797} (\bibinfo {year} {1989})}\BibitemShut {NoStop}%
\bibitem [{\citenamefont {Nikoli\ifmmode~\acute{c}\else
  \'{c}\fi{}}(2020)}]{Nikolic2020}%
  \BibitemOpen
  \bibfield  {author} {\bibinfo {author} {\bibfnamefont {P.}~\bibnamefont
  {Nikoli\ifmmode~\acute{c}\else \'{c}\fi{}}},\ }\bibfield  {title} {\bibinfo
  {title} {Quantum field theory of topological spin dynamics},\ }\href
  {https://doi.org/10.1103/PhysRevB.102.075131} {\bibfield  {journal} {\bibinfo
   {journal} {Phys. Rev. B}\ }\textbf {\bibinfo {volume} {102}},\ \bibinfo
  {pages} {075131} (\bibinfo {year} {2020})}\BibitemShut {NoStop}%
\bibitem [{\citenamefont {Pradenas}\ and\ \citenamefont
  {Tchernyshyov}(2024)}]{Tchernyshyov_2024}%
  \BibitemOpen
  \bibfield  {author} {\bibinfo {author} {\bibfnamefont {B.}~\bibnamefont
  {Pradenas}}\ and\ \bibinfo {author} {\bibfnamefont {O.}~\bibnamefont
  {Tchernyshyov}},\ }\bibfield  {title} {\bibinfo {title} {Spin-frame field
  theory of a three-sublattice antiferromagnet},\ }\href
  {https://doi.org/10.1103/PhysRevLett.132.096703} {\bibfield  {journal}
  {\bibinfo  {journal} {Phys. Rev. Lett.}\ }\textbf {\bibinfo {volume} {132}},\
  \bibinfo {pages} {096703} (\bibinfo {year} {2024})}\BibitemShut {NoStop}%
\bibitem [{\citenamefont {Watanabe}\ and\ \citenamefont
  {Vishwanath}(2014)}]{Watanabe_2014}%
  \BibitemOpen
  \bibfield  {author} {\bibinfo {author} {\bibfnamefont {H.}~\bibnamefont
  {Watanabe}}\ and\ \bibinfo {author} {\bibfnamefont {A.}~\bibnamefont
  {Vishwanath}},\ }\bibfield  {title} {\bibinfo {title} {Criterion for
  stability of goldstone modes and fermi liquid behavior in a metal with broken
  symmetry},\ }\href {https://doi.org/10.1073/pnas.1415592111} {\bibfield
  {journal} {\bibinfo  {journal} {Proceedings of the National Academy of
  Sciences}\ }\textbf {\bibinfo {volume} {111}},\ \bibinfo {pages}
  {16314–16318} (\bibinfo {year} {2014})}\BibitemShut {NoStop}%
\bibitem [{\citenamefont {Watanabe}\ \emph {et~al.}(2014)\citenamefont
  {Watanabe}, \citenamefont {Parameswaran}, \citenamefont {Raghu},\ and\
  \citenamefont {Vishwanath}}]{Watanabe_2014_Skx}%
  \BibitemOpen
  \bibfield  {author} {\bibinfo {author} {\bibfnamefont {H.}~\bibnamefont
  {Watanabe}}, \bibinfo {author} {\bibfnamefont {S.~A.}\ \bibnamefont
  {Parameswaran}}, \bibinfo {author} {\bibfnamefont {S.}~\bibnamefont
  {Raghu}},\ and\ \bibinfo {author} {\bibfnamefont {A.}~\bibnamefont
  {Vishwanath}},\ }\bibfield  {title} {\bibinfo {title} {Anomalous fermi-liquid
  phase in metallic skyrmion crystals},\ }\href
  {https://doi.org/10.1103/PhysRevB.90.045145} {\bibfield  {journal} {\bibinfo
  {journal} {Phys. Rev. B}\ }\textbf {\bibinfo {volume} {90}},\ \bibinfo
  {pages} {045145} (\bibinfo {year} {2014})}\BibitemShut {NoStop}%
\bibitem [{\citenamefont {Watanabe}(2020)}]{Watanabe_2020}%
  \BibitemOpen
  \bibfield  {author} {\bibinfo {author} {\bibfnamefont {H.}~\bibnamefont
  {Watanabe}},\ }\bibfield  {title} {\bibinfo {title} {Counting rules of
  nambu–goldstone modes},\ }\href
  {https://doi.org/10.1146/annurev-conmatphys-031119-050644} {\bibfield
  {journal} {\bibinfo  {journal} {Annual Review of Condensed Matter Physics}\
  }\textbf {\bibinfo {volume} {11}},\ \bibinfo {pages} {169–187} (\bibinfo
  {year} {2020})}\BibitemShut {NoStop}%
\bibitem [{\citenamefont {Cr\'epel}\ \emph {et~al.}(2023)\citenamefont
  {Cr\'epel}, \citenamefont {Guerci}, \citenamefont {Cano}, \citenamefont
  {Pixley},\ and\ \citenamefont {Millis}}]{Crepel_2023}%
  \BibitemOpen
  \bibfield  {author} {\bibinfo {author} {\bibfnamefont {V.}~\bibnamefont
  {Cr\'epel}}, \bibinfo {author} {\bibfnamefont {D.}~\bibnamefont {Guerci}},
  \bibinfo {author} {\bibfnamefont {J.}~\bibnamefont {Cano}}, \bibinfo {author}
  {\bibfnamefont {J.~H.}\ \bibnamefont {Pixley}},\ and\ \bibinfo {author}
  {\bibfnamefont {A.}~\bibnamefont {Millis}},\ }\bibfield  {title} {\bibinfo
  {title} {Topological superconductivity in doped magnetic moir\'e
  semiconductors},\ }\href {https://doi.org/10.1103/PhysRevLett.131.056001}
  {\bibfield  {journal} {\bibinfo  {journal} {Phys. Rev. Lett.}\ }\textbf
  {\bibinfo {volume} {131}},\ \bibinfo {pages} {056001} (\bibinfo {year}
  {2023})}\BibitemShut {NoStop}%
\bibitem [{\citenamefont {{Doniach}}(1977)}]{Doniach_1977}%
  \BibitemOpen
  \bibfield  {author} {\bibinfo {author} {\bibfnamefont {S.}~\bibnamefont
  {{Doniach}}},\ }\bibfield  {title} {\bibinfo {title} {{The Kondo lattice and
  weak antiferromagnetism}},\ }\href
  {https://doi.org/10.1016/0378-4363(77)90190-5} {\bibfield  {journal}
  {\bibinfo  {journal} {Physica B+C}\ }\textbf {\bibinfo {volume} {91}},\
  \bibinfo {pages} {231} (\bibinfo {year} {1977})}\BibitemShut {NoStop}%
\bibitem [{\citenamefont {Evers}\ and\ \citenamefont
  {Mirlin}(2008)}]{RevModPhys.80.1355}%
  \BibitemOpen
  \bibfield  {author} {\bibinfo {author} {\bibfnamefont {F.}~\bibnamefont
  {Evers}}\ and\ \bibinfo {author} {\bibfnamefont {A.~D.}\ \bibnamefont
  {Mirlin}},\ }\bibfield  {title} {\bibinfo {title} {Anderson transitions},\
  }\href {https://doi.org/10.1103/RevModPhys.80.1355} {\bibfield  {journal}
  {\bibinfo  {journal} {Rev. Mod. Phys.}\ }\textbf {\bibinfo {volume} {80}},\
  \bibinfo {pages} {1355} (\bibinfo {year} {2008})}\BibitemShut {NoStop}%
\end{thebibliography}%


\onecolumngrid
\newpage
\makeatletter 

\begin{center}
\textbf{\large Supplementary materials for: ``\@title ''} \\[10pt]
Daniele Guerci$^{1,2}$, J. H. Pixley$^{3,2}$ and Andrew Millis$^{3,4}$ \\
\textit{$^1$ Department of Physics, Massachusetts Institute of Technology, Cambridge, MA, USA}\\
\textit{$^2$ Center for Computational Quantum Physics, Flatiron Institute, 162 5th Avenue, NY 10010, USA}\\
\textit{$^3$ Department of Physics and Astronomy, Center for Materials Theory, Rutgers University, Piscataway, New Jersey 08854, USA}\\
\textit{$^4$Department of Physics, Columbia University, 538 Wst 120th Street, New York, NY 10027}
\end{center}
\vspace{20pt}

\setcounter{figure}{0}
\setcounter{section}{0}
\setcounter{equation}{0}

\renewcommand{\thefigure}{S\@arabic\c@figure}
\makeatother

\appendix 

\section{Conventions}
	\label{app:conventions} 
	
We choose a system of coordinates in which the scalar product of the $\pm {\bm \kappa}$ vectors with the nearest neighbor vectors ${\bm u}_{j=1,2,3}$ and ${\bm \gamma}_{j=1,2,3}$ are given by
	\begin{equation}
		\label{convention}
		\begin{array}{|c||c|c|c|c|c|c|} 
			{\bm r} & {\bm \gamma}_1 &  {\bm \gamma}_2 &  {\bm \gamma}_3 &  {\bm u }_1 & {\bm u}_2 &  {\bm u}_3 \\\hline\hline
			e^{i\bm \kappa \cdot {\bm r}} & \omega^* & \omega^* & \omega^* & 1 & \omega & \omega^* \\\hline
			e^{i\bm\kappa' \cdot {\bm r}} & \omega & \omega & \omega & 1 & \omega^* &  \omega 
		\end{array} \, \, , \quad \omega = e^{2i\pi/3} .
	\end{equation}
This can be realized with ${\bm u}_j =C^{j-1}_{3z}{\bm u}_1$ and ${\bm u}_1=a(1,0)/\sqrt{3}$, $\bm \kappa= -\bm\kappa' = (4\pi/3a)(0,1)$ and ${\bm \gamma}_1=a(0,1)$, ${\bm \gamma}_j=C^{j-1}_{3z}{\bm\gamma}_1$.

\section{Minimal model for AB-stacked MoTe$_2$/WSe$_2$}

The Hamiltonian describing the properties of AB-stacked MoTe$_2$/WSe$_2$ reads:
\begin{equation}
\label{H_tb}
\begin{split}
    H_{\rm int}=&-\frac{\Delta}{2} (N_{\rm M}-N_{\rm W})+U_{\mathrm{M}}\sum_{\br\in {\rm M}}n_{\br\uparrow}n_{\br\downarrow}+U_{\mathrm{W}}\sum_{\br\in W}n_{\br\uparrow}n_{\br\downarrow}\\
    H_0=&- t_{\rm W}\sum_{\langle \br,\br'\rangle\in W} e^{-i\nu_{\br,\br'}\varphi s_\sigma} c^\dagger_{\br\sigma} c_{\br'\sigma}-t_{\rm M} \sum_{\langle \br,\br'\rangle\in {\rm M}} f^\dagger_{\br\sigma} f_{\br'\sigma}-t_\perp\sum_{\langle\br,\br'\rangle}f^\dagger_{\br}c_{\br'}.
\end{split}
\end{equation}
This Hamiltonian can be derived directly from Wannierization of the low energy orbitals of the continuum model in Ref.~\cite{Zhang_2021,prb_daniele_2024}. The interplay of the spin orbit coupling and the opposite angular momentum of the $d$-shell orbitals residing in the two layers leads to a topological non-trivial interlayer hopping.

\subsection{Symmetries of the lattice model}
\label{app:symmetries}

The symmetries of the lattice model are the spin U(1) $e^{i S^z\theta}$ with $S^z$ spin operator, threefold rotation symmetry around z $C_{3z}$, the mirror $\mathcal M_y$ and the time reversal symmetry $\mathcal T$. $C_{3z}$ acts on the Fermi field operators as: 
\begin{equation}
\label{threefold_lattice}
 C_{3z} c_{\br\sigma} C^\dagger_{3z}=c_{C_{3z}\br\sigma},\quad  C_{3z} f_{\br\sigma} C^\dagger_{3z}=f_{C_{3z}\br\sigma},
\end{equation}
corresponding to a spin-space three-fold rotation. $\mathcal M_y$ is the combination of a mirror symmetry $y\to-y$ and a spin/valley flip $\uparrow\leftrightarrow\downarrow$
\begin{equation}
     \mathcal M_y c_{\br\sigma} \mathcal M_y=\sigma^x_{\sigma\sigma'}c_{\mathcal M_y\br\sigma'},\quad \mathcal M_y f_{\br\sigma} \mathcal M_y=\sigma^x_{\sigma\sigma'}f_{\mathcal M_y\br\sigma'}. 
\end{equation}
Finally, we have the time-reversal symmetry $\mathcal{T}=i\sigma^y \mathcal{K}$:
\begin{equation}
    \mathcal{T} c_{\br\sigma} \mathcal{T}^{-1}=(-i\sigma^y)_{\sigma\sigma'}c_{\br\sigma'},\quad \mathcal{T} f_{\br\sigma} \mathcal{T}^{-1}=(-i\sigma^y)_{\sigma\sigma'}f_{\br\sigma'}.
\end{equation}
Table~\ref{tab:symmetries_states} includes phases realized at filling $1+x$ with $x>0$ in our tight-binding model for AB-stacked heterobilayer and their symmetries. 
\begin{table}[]
\centering
\begin{tabular}{|c||c|c|c|c|c|}
\hline
 Phase & $C_{3z}$ & $\mathcal M_y$ & $S^z$ & $\mathcal T$ & $\Theta=S^z\mathcal T$    \\
\hline\hline
AFM & \cmark & \cmark & \xmark  & \xmark & \cmark  \\
\hline
cAFM & \cmark & \xmark & \xmark & \xmark & \xmark   \\
\hline
FM$_z$ & \cmark & \xmark & \cmark & \xmark & \xmark  \\
\hline
hFL & \cmark & \cmark & \cmark & \cmark & \cmark  \\
\hline
\end{tabular}
\caption{Different rows correspond to different phases including the in-plane and the canted 120$^\circ$ N\'eel order denoted as AFM and cAFM, respectively. FM$_z$ and hFL refer to the out of plane ferromagnet and the paramagnetic heavy Fermi liquid. Different columns show the symmetries of the lattice model including the threefold rotational symmetry $C_{3z}$, mirror $\mathcal M_y$, U(1) spin/valley generated by $S^z$, time reversal symmetry $\mathcal T$ and $\Theta$. 
} 
\label{tab:symmetries_states}
\end{table}

\subsection{The tetrahedron model}
\label{app:tetra}

\begin{figure}
    \centering
    \includegraphics[width=0.7\linewidth]{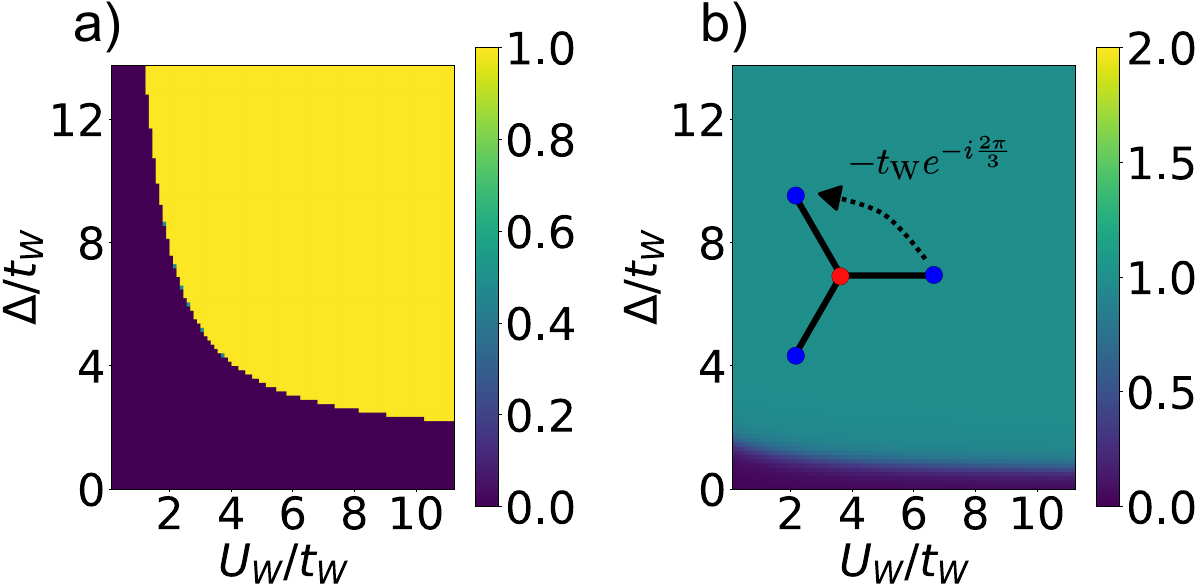}
        \caption{Phase diagram of the two electron sector $(Q=2)$ of the tetrahedron model composed of the central site corresponding to the local moment and the three neighboring sites as shown in the inset of panel b). a) Average value of $\left(S^{z}_{\rm tot}\right)^2$ for the ground state in the sector $Q=2$ with $S^{z}_{\rm tot}$ total spin. 
        b) Average value of $n_f$ taking value $1$ in the local moment regime (dark green region). Strong valence fluctuations are present in the small and large $\Delta$ regime where the local moment description is not applicable. Calculations are performed setting $t_{\rm W}=8$meV, $t_\perp=2$meV, $U_{\rm M}=100$meV and phase $\varphi=2\pi/3$.}
    \label{fig:tetrahedron}
\end{figure}

The origin of magnetic correlations are understood from a minimal model composed by four sites, two carrier ``tetrahedron'' model shown in the inset of Fig.~\ref{fig:tetrahedron}b). Though simple, the model incorporating charge fluctuations within the elementary plaquette effectively captures the spin-exchange mechanism that drives easy-axis anisotropy and ferromagnetism.  This model includes the central local moment site belonging to Mo and three adjacent sites describing the bottom itinerant layer, and has the $t_W$, $t_\perp$ $\Delta$ and $U_l$:
\begin{equation}
    H_{\rm int}=-\frac{\Delta}{2}\left(n_f-\sum_{j=1}^{3} n_{c j}\right)+U_{\mathrm{M}}n_{f\uparrow}n_{f\downarrow}+U_{\mathrm{W}}\sum_{j=1}^{3}n_{c j\uparrow}n_{c j\downarrow},
\end{equation}
and it also includes the hopping 
\begin{equation}
    H_{0} = -t_{\rm W}\sum_{j=1}^{3}\sum_{\sigma}\left[c^\dagger_{j+1\sigma}c_{j\sigma} e^{-i\varphi\sigma}+h.c.\right] - t_\perp\sum_{j=1}^{3} \sum_{\sigma}\left[ f^\dagger_\sigma c_{j\sigma}+h.c.\right]
\end{equation}
where in the following we set $\varphi=2\pi/3$. While the spin-orbit coupling means the total spin is not a good quantum number, the $z$ component of the total spin, $S^{z}_{\rm tot}$ commutes with the Hamiltonian.
Numerical diagonalization is performed in the charge sector $Q=2$ in 4 sites exploring the different spin sectors $S^z=0,\pm 1$ composed in total by 28 states. 
Panel (a) of Fig.~\ref{fig:tetrahedron} shows $\langle \left(S^{z}_{\rm tot}\right)^2\rangle$ of the plaquette as a function of $\Delta$ and $U_{\rm W}$ for large $U_{\rm M}\approx 12t_{\rm W}$. High spin ($\langle  \left(S^{z}_{\rm tot}\right)^2\rangle=1$) and low spin ($\langle \left(S^{z}_{\rm tot}\right)^2\rangle=0$ states) are seen. 
The low spin state favored at $U_W=0$ undergoes a transition to a high spin state at a $\Delta$-dependent value of $U_W$. 
This transition highlights the previously un-noticed role of the itinerant carrier interaction in producing a {\it ferromagnetic} Kondo exchange between the local moment and the itinerant electron. 
Panel (b) of Fig.~\ref{fig:tetrahedron} shows that the ferromagnetic correlations (high spin state) occur in the ``Kondo'' regime in which $n_f\approx 1$.

\section{Spin-fermion model Hamiltonian: classical solution}
\label{app:classical_limit}

In this Appendix we discuss the solution of the spin-fermion Hamiltonian:
\begin{equation}
    H_{S} = - t_{\rm W}\sum_{\langle \br,\br'\rangle\in \rm W} e^{-i\nu_{\br,\br'}\varphi s_\sigma} c^\dagger_{\br\sigma}c_{\br'\sigma} + J_H\sum_{\langle\br,\br'\rangle}\bm S_{\br}\cdot\bm S_{\br'} + U_{W}\sum_{\br \in {\rm W}}  n_{\br \uparrow} n_{\br \downarrow} + H_K, 
\end{equation}
in the classical limit where we replace the spin with its expectation value $\bm S_{\br}\to \bm M$. Additionally, we discuss the role of fluctuations around the classical solution. In the following we will focus our discussion to the case $\varphi=2\pi/3$.

\subsection{Itinerant carriers in a classical background: canted 120$^\circ$ antiferromagnet}

\begin{figure}
    \centering
    \includegraphics[width=0.6\linewidth]{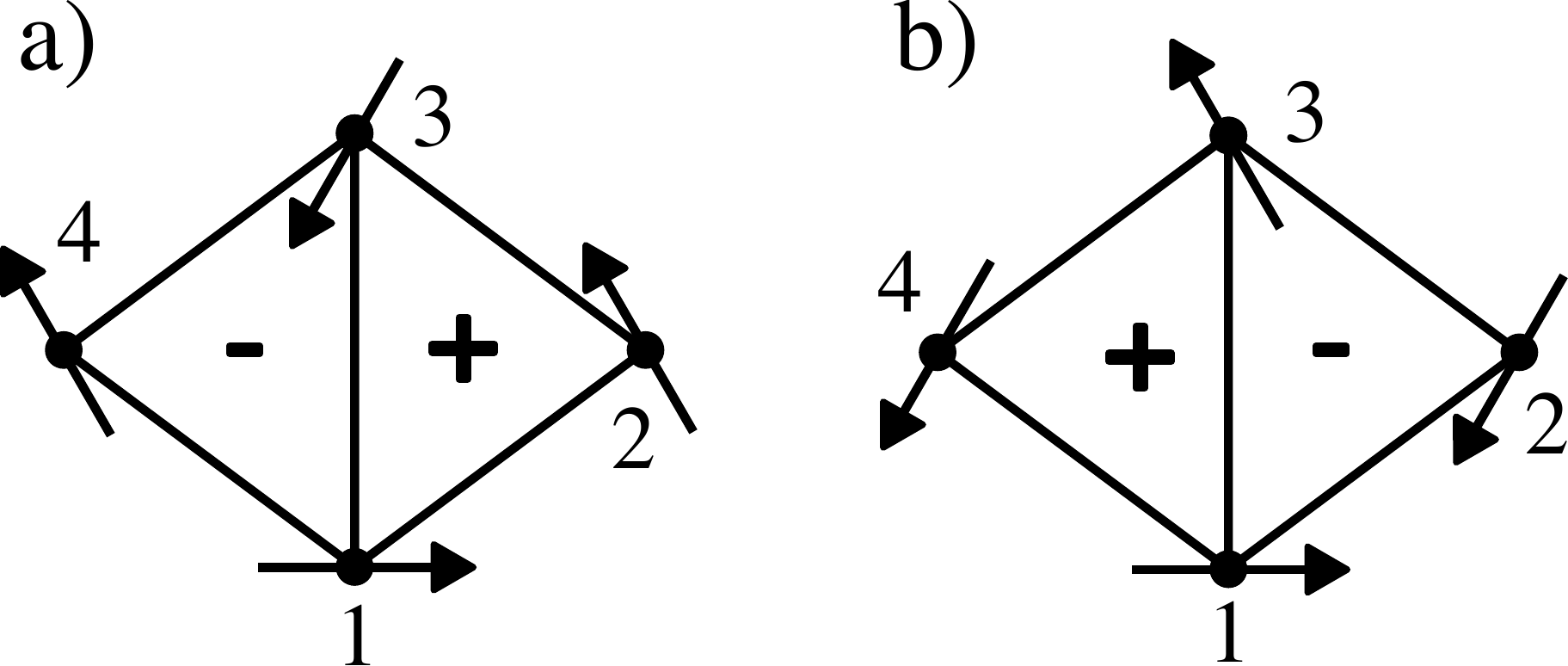}
    \caption{Sketch of the 120$^\circ$ N\'eel order for wavevector modulations $\bm Q=\bm\kappa'-\bm\kappa=\bm q_1$ a) and $\bm Q=\bm\kappa-\bm\kappa'=-\bm q_1$ b). The two different modulations are characterized by opposite winding number $\chi=\pm$~\eqref{chirality}.  }
    \label{fig:sketch_neel_chirality}
\end{figure}

In this section we simply replace the spin degree of freedom with a classical vector satisfying $\bm S_{\bm r}\cdot\bm S_{\bm r}=1/4$. In addition to the in-plane 120$^\circ$ antiferromagnetic we include an out-of-plane spin canting $M_z$: 
\begin{equation}
    \bm S_{\br}=M_z\bm z+\left(\bm S_{\bm Q}e^{i\bm Q\cdot\bm r}+\bm S_{-\bm Q}e^{-i\bm Q\cdot\bm r}\right),
\end{equation}
where $\bm S_{\bm Q}=M_\parallel(\bm x-i\bm y)/2$ and $M^2_z+M^2_\parallel=1/4$ implying $M_z=\sin(\theta)/2$ and $M_\parallel=\cos(\theta)/2$ with $\theta$ canting angle. Replacing the quantum spin operator with its classical expectation value we find 
\begin{equation}\label{H_classical_1}
    \begin{split}
    H_S[\{\bm S_{\br}\}]&=-\frac{3}{4}NJ_H\cos\theta^2+\frac{3}{2}NJ_H\sin\theta^2+U_{\rm W}\sum_{\br}n_{\bm r\uparrow}n_{\bm r\downarrow} - t_{\rm W}\sum_{\langle \br,\br'\rangle\in \rm W} e^{-i\nu_{\br,\br'}2\pi s_\sigma/3} c^\dagger_{\br\sigma}c_{\br'\sigma} \\
    & + \frac{\sin\theta}{4}\sum_{\br}\left[J^{(I)}_K\sum^{i\neq j}_{ij}c^\dagger_{\br+\bm u_i}\sigma^zc_{\br+\bm u_j}+J^{(II)}_K\sum_{i}c^\dagger_{\br+\bm u_i}\sigma^zc_{\br+\bm u_i}\right]\\
    &+ \frac{\cos\theta}{4}\sum_{\br}e^{i\bm Q\cdot\br}\left[J^{(I)}_K\sum^{i\neq j}_{ij}c^\dagger_{\br+\bm u_i}\left( e^{i\phi}\sigma^-\right)c_{\br+\bm u_j}+J^{(II)}_K\sum_{i}c^\dagger_{\br+\bm u_i}\left( e^{i\phi}\sigma^-\right)c_{\br+\bm u_i}\right]+h.c.\\
    &= E_S[\theta,\phi] + \mathcal H_\psi[\theta,\phi], 
    \end{split}
\end{equation}
where $\theta$, $\phi$ are variational parameters.
Notice that the dependency of the energy on $\phi$ can be gauge out by performing the transformation $c_{\uparrow}\to c_{\uparrow}e^{-i\phi}$. Therefore, the energy does not depend on the in-plane orientation of the 120$^\circ$ order. Fig.~\ref{fig:sketch_neel_chirality}a) and b) shows two different 120$^\circ$ ordering with opposite winding number $\chi=\pm$, defined as: 
\begin{equation}\label{chirality}
    \chi=2\bm z\cdot(\bm S_1\times\bm S_2+\bm S_2\times\bm S_3+\bm S_3\times\bm S_1)/(3\sqrt{3}).
\end{equation}
Interestingly, the coupling with the itinerant electrons breaks the mirror $x$ symmetry $\mathcal M_x$ and differentiates states with opposite chirality $\chi=\pm$ depending on the ordering wavevector taking these two signs $\bm Q_{\mathrm{or}}=\pm(\bm\kappa'-\bm\kappa)$. 
In the dilute regime, away from the Van Hove singularity, the on-site interaction renormalizes the properties of the electrons in the WSe$_2$ layer. The simplest approach to include the renormalization induced by the interaction is the slave boson approach  which consists of replacing $c_{\br\sigma}$ with: 
\begin{equation}
    c_{\br\sigma}=b^\dagger_{\br}\psi_{\br\sigma},\quad     b^\dagger_{\bm r} b_{\bm r}+\sum_\sigma\psi^\dagger_{\bm r\sigma}\psi_{\bm r\sigma}=1,
\end{equation}
the latter condition projecting to the physical Hilbert space which in the large $U_{\rm W}$ limit includes only holons and singly occupied sites. Within the mean-field approximation we find $\langle b_{\br}\rangle=b_0=1-x$ is condensed and the effective Hamiltonian describing the fermions reads:  
\begin{equation}\label{H_classical_2}
    \begin{split}
     \mathcal H_{\rm \psi}[\theta,\phi]&=- t_{\rm W}|b_0|^2\sum_{\langle \br,\br'\rangle\in \rm W} e^{-i\nu_{\br,\br'}2\pi s_\sigma/3} \psi^\dagger_{\br\sigma}\psi_{\br'\sigma} \\
    & + \frac{\sin\theta}{4}\sum_{\br}\left[J^{(I)}_K|b_0|^2\sum^{i\neq j}_{ij}\psi^\dagger_{\br+\bm u_i}\sigma^z\psi_{\br+\bm u_j}+J^{(II)}_K\sum_{i}\psi^\dagger_{\br+\bm u_i}\sigma^z\psi_{\br+\bm u_i}\right]\\
    &+ \frac{\cos\theta}{4}\sum_{\br}e^{i\bm Q_{\rm or}\cdot\br}\left[J^{(I)}_K|b_0|^2\sum^{i\neq j}_{ij} \psi^\dagger_{\br+\bm u_i} \sigma^-\psi_{\br+\bm u_j}+J^{(II)}_K\sum_{i}\psi^\dagger_{\br+\bm u_i}\sigma^-\psi_{\br+\bm u_i}\right]+h.c.
    \end{split}
\end{equation}
In the plane wave basis we have the Hartree-Fock Hamiltonian: 
\begin{equation}\label{H_classical_3}
    \begin{split}
     \mathcal H_{\rm \psi}[\theta,\phi]=\sum_{\bk}\bar\epsilon_{\bk\sigma}\psi^\dagger_{\bk\sigma}\psi_{\bk\sigma}+\frac{\sin\theta}{4}\sum_{\bk}J_{\bk,\bk}\psi^\dagger_{\bk}\sigma^z\psi_{\bk}+\frac{\cos\theta}{4}\sum_{\bk}\left[J_{\bk+\bm Q_{\rm or},\bk}\psi^\dagger_{\bk+\bm Q_{\rm or}}\sigma^-\psi_{\bk}+h.c.\right],
    \end{split}
\end{equation}
where $\bm Q_{\rm or}$ is the wave vector modulation of the 120$^\circ$ N\'eel state and we have introduced the momentum dependent exchange: 
\begin{equation}
    J_{\bk,\bp}=J^{(I)}_{K}|b_0|^2\left(V^*_{\bk}V_{\bp}-V_{\bp-\bk}\right)+J^{(II)}_K V_{\bp-\bk},\quad V_{\bk}=\sum^3_{j=1}e^{i\bk\cdot\bm u_j},
\end{equation} 
and the dispersion 
\begin{equation}
    \bar\epsilon_{\bk\sigma}=-2t_{\rm W}|b_0|^2\sum^3_{j=1}\cos(\bk\cdot\bm \gamma_j+2\pi\sigma/3).
\end{equation}
In the latter expression we have introduced $\bm\gamma_j=a\omega^{j-1}$ with $\omega=\exp(2\pi i/3)$ in complex notation. 
In the reduced Brillouin zone $(\tilde{\bm b}_1,\tilde{\bm b}_2)$ with $\tilde{\bm b}_{1/2}=\frac{4\pi}{3a}(\pm \sqrt{3}/2,1/2)$ the Hamiltonian $h_{\chi=\pm}$~\eqref{H_classical_3} with $\chi=\pm$ associated to the two chiralities $\pm\bm Q$ reads: 
\begin{equation}\label{h_plus_minus}
\begin{split}
    &h_{+}(\bk) = \begin{pmatrix}
        \bar h_{\bk+\bm q_1} + \frac{\sin\theta}{4}J_{\bk+\bm q_1,\bk+\bm q_1}\sigma^z & \frac{\cos\theta}{4} J_{\bk+\bm q_1,\bk}\sigma^- & h.c. \\ 
        h.c. & \bar h_{\bk} + \frac{\sin\theta}{4}J_{\bk,\bk}\sigma^z & \frac{\cos\theta}{4} J_{\bk,\bk-\bm q_1}\sigma^- \\ 
        \frac{\cos\theta}{4} J_{\bk-\bm q_1,\bk+\bm q_1}\sigma^- & h.c. & \bar h_{\bk-\bm q_1} + \frac{\sin\theta}{4}J_{\bk-\bm q_1,\bk-\bm q_1}\sigma^z 
    \end{pmatrix},\\
    &h_{-}(\bk) = \begin{pmatrix}
        \bar h_{\bk+\bm q_1} + \frac{\sin\theta}{4}J_{\bk+\bm q_1,\bk+\bm q_1}\sigma^z & h.c. & \frac{\cos\theta}{4}J_{\bk+\bm q_1,\bk-\bm q_1}\sigma^- \\ 
        \frac{\cos\theta}{4} J_{\bk,\bk+\bm q_1}\sigma^-  & \bar h_{\bk} + \frac{\sin\theta}{4}J_{\bk,\bk}\sigma^z & h.c. \\ 
        h.c. & \frac{\cos\theta}{4} J_{\bk-\bm q_1,\bk}\sigma^-  & \bar h_{\bk-\bm q_1} + \frac{\sin\theta}{4}J_{\bk-\bm q_1,\bk-\bm q_1}\sigma^z 
    \end{pmatrix},
\end{split}
\end{equation}
where $\bar h_{\bk}=\text{diag}\left[\bar\epsilon_{\bk\uparrow},\bar\epsilon_{\bk\downarrow}\right]$. Finally, we find that the classical energy of the spin texture reads: 
\begin{equation}
    E[\{\bm S_{\br}\}]=E_S+\mel{\Psi_{\psi}}{\mathcal H_{\psi}}{\Psi_\psi},
\end{equation}
where $\ket{\Psi_{\psi}}$ is the Slater determinant describing the ground state of the Hamiltonian~\eqref{H_classical_3}.

\subsection{The low-energy limit}

\begin{figure}
    \centering
    \includegraphics[width=1\linewidth]{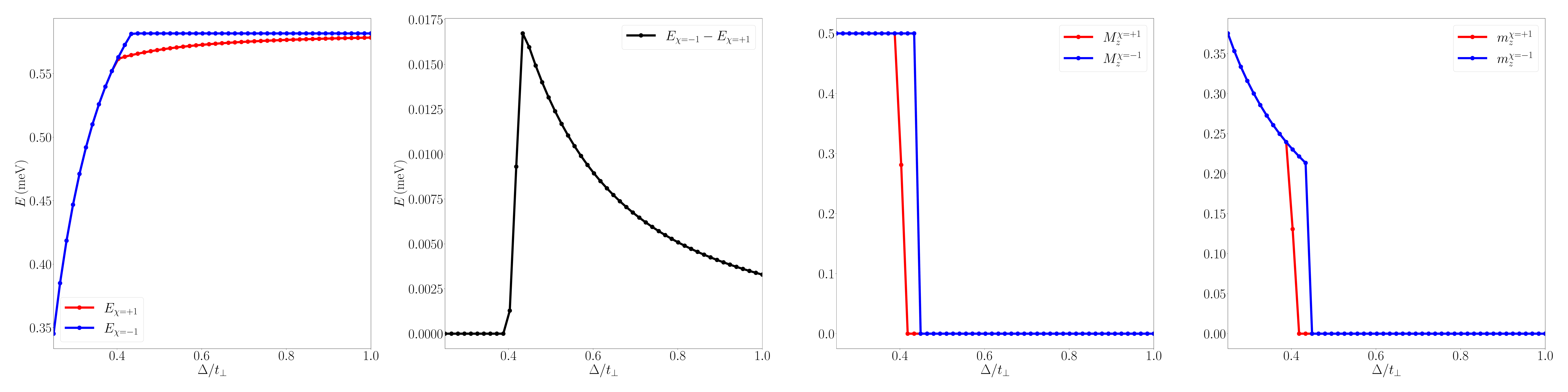}
    \caption{Energy and out of plane magnetization of the two competing magnetic ground state. $\chi=+1$ is energetically favorable at any finite $M_\parallel$. For $M_\parallel=0$ the ground state is a fully polarized ferromagnet in both cases. The calculation has been performed setting $x=0.15$, $t_{\rm W}=7$meV, $J_H=0.1$meV, $t_\perp=2$meV, $U_{\rm W}=70$meV, $U_{\rm M}=100$meV.
    }
    \label{fig:comparison_chi}
\end{figure}

To provide analytical results we perform an expansion around $\kappa$ and $\kappa'$ where the Fermi surface of the low-density hole gas is located. Before moving on, we provide the general expansions of the form factors in the Hamiltonian~\eqref{h_plus_minus} around $\bm\kappa$ and $\bm\kappa'$.  
\begin{equation}
    V_{\bm \kappa+\bp}\approx \frac{i\sqrt{3}}{2}p_+,\quad    V_{\bm \kappa'+\bp}\approx \frac{i\sqrt{3}}{2}p_-.
\end{equation}
In second order in $\bk$ we have: 
\begin{equation}
    J_{\bk+\bm \kappa,\bk+\bm \kappa}\approx-3\left[J^{(I)}_K|b_0|^2-J^{(II)}_K\right]+\frac{3J^{(I)}_K|b_0|^2}{4} k^2,
\end{equation}
and 
\begin{equation}
    J_{\bk+\bm \kappa',\bk+\bm \kappa}\approx\frac{3J^{(I)}_K|b_0|^2}{4} k^2_+,\quad 
    J_{\bk+\bm \kappa',\bk+\bm \kappa}\approx\frac{3J^{(I)}_K|b_0|^2}{4} k^2_-. 
\end{equation}
 We introduce the two coupling constants: 
\begin{equation}
    J_z =3\left[J^{(I)}_K|b_0|^2-J^{(II)}_K\right],\quad J_\perp=3J^{(I)}_K|b_0|^2/4.
\end{equation}
In the following, we consider separately the cases $\chi=\pm$ which are characterized by a different low-energy Hamiltonian. We anticipate that the $\chi=+$ state is always energetically favorable as shown in Fig.~\ref{fig:comparison_chi} for filling factor $x=0.15$.

\subsubsection{Winding number $\chi=-$}

Introducing the notation $\Psi_{\bk\Uparrow}\equiv \psi_{\bm\kappa+\bk\uparrow}$ and $\Psi_{\bk\Downarrow}\equiv \psi_{\bm\kappa'+\bk\downarrow}$ for $|\bk a|\ll1$ we find: 
\begin{equation}\label{H_classical_4}
h_-(\bk) \approx \frac{\bm k^2}{2m^*_{\rm W}} \sigma^0-\left[\frac{J_z M_z}{2}-\frac{J_\perp M_z \bk^2 }{2}\right]\sigma^z. 
\end{equation}
Given $M_z$ and the filling factor $x$ the chemical potential are $\mu_\pm=\min[(\mu\pm J_z M_z/2),0]/(1\pm J_\perp m^*_{\rm W}M_z)$. The value of $\mu$ is obtained as $x=\rho^*_{\rm W}\left[\mu_++\mu_-\right]$. Then energy density is given by: 
\begin{equation}
    \mathcal  E^{\chi=-}_{\rm cl}=-\frac{3 J_HM^2_\parallel}{2}+3J_ H M^2_z-J_zM_z\frac{\rho_{\rm W}^*(\mu_+-\mu_-)}{2}+\frac{\rho^*_{\rm W}}{2}\sum_{s=\pm}(1+s m^*_{\rm W} J_\perp M_z )\mu^2_s. 
\end{equation}
To determine the critical value of the coupling $J_z$ and $J_\perp$ we expand the latter expression for small $M_z$.

\subsubsection{Winding number $\chi=+$}

Performing the $\bk\cdot\bp$ expansion of $h_+$ and keeping only the low-energy sector we find the Hamiltonian in the basis $[\Psi_{\bk\Uparrow},\Psi_{\bk\Downarrow}]^T$:
\begin{equation}\label{H_classical_5}
    h_+(\bk) \approx \frac{\bk^2}{2m^*_{\rm W}}\sigma^0-J_zM_z\frac{\sigma^z}{2}+J_\perp M_z\, \bk^2\frac{\sigma^z}{2}+J_\perp M_\parallel\frac{k^2_+\sigma^-+h.c.}{2},
\end{equation}
We notice that the itinerant carrier spin is not a good quantum number due to the effective spin orbit coupling (SOC) originating from the N\'eel order. The SOC introduces a vortex at $\bk=0$ with Berry phase $2\pi$. Similarly to the previous case, the value of the chemical potential is determined as $x=\rho^*_{\rm W}[\mu_++\mu_-]$ where: 
\begin{equation}
    \mu_\pm = \frac{\mu-\left(J_zJ_\perp m^*_{\rm W} M^2_z\right)/2}{1-(J_\perp m^*_{\rm W}/2)^2}\mp\sqrt{\left[\frac{\mu-J_zJ_\perp M^2_zm^*_{\rm W}/2}{1-(J_\perp m^*_{\rm W}/2)^2}\right]^2-\frac{\mu^2-(J_z M_z/2)^2}{1-(J_\perp m^*_{\rm W}/2)^2}},
\end{equation}
$\mu_->\mu_+$. Introducing the vectors $\bm d_{\bk} = \left[J_\perp M_\parallel(k^2_x-k^2_y)/2,J_\perp M_\parallel2k_xk_y/2,-J_zM_z/2+J_\perp M_z\bk^2/2\right]$, the eigenstates of the Hamiltonian are the spinor: 
\begin{equation}
   \ket{ \bm \chi_{\bk+}}=\begin{pmatrix}
        \cos\theta_{\bk}/2\\
        e^{i\varphi_{\bk}}\sin\theta_{\bk}/2
    \end{pmatrix},\quad    \ket{ \bm \chi_{\bk-}}=\begin{pmatrix}
        -\sin\theta_{\bk}/2\\
        e^{i\varphi_{\bk}}\cos\theta_{\bk}/2
    \end{pmatrix} 
\end{equation}
where the corresponding energy is $E_{\bk\pm}=\bk^2/2m^*_{\rm W}\pm |\bm d_{\bk}|$, $\cos \theta_{\bk}=d_z/|\bm d|$ and $\tan \varphi_{\bk}=d_y/d_x$. By definition we have: 
\begin{equation}
    \mel{\bm \chi_{\bk\pm }}{\bm \sigma }{\bm \chi_{\bk\pm }}=\pm \bm n_{\bk} =\pm \bm d_{\bk}/|\bm d_{\bk}|,
\end{equation}
and 
 \begin{equation}
     \langle\Psi^\dagger_{\bk}\sigma^a\Psi_{\bk}\rangle= n^a_{\bk}\left[f(E_{\bk+}-\mu)-f(E_{\bk-}-\mu)\right].
 \end{equation}
 The energy becomes: 
 \begin{equation}
    \begin{split}
    \mathcal  E^{\chi=+}_{\rm cl}&=-\frac{3 J_HM^2_\parallel}{2}+3J_ H M^2_z+\frac{\rho^*_{\rm W}(\mu^2_++\mu^2_-)}{2}-\rho^*_{\rm W}\left(m^*_{\rm W} M_\parallel J_\perp\right)^2\int^{\mu_-}_{\mu_+}{d\epsilon}\frac{\epsilon^2}{|\bm d_\epsilon|} \\
    &-\rho^*_{\rm W}\frac{J_z M_z}{2}\int^{\mu_-}_{\mu_+}{d\epsilon}\frac{J_z M_z-2m^*_{\rm W}\epsilon J_\perp M_z}{2|\bm d_\epsilon|}+\rho^*_{\rm W}\frac{J_\perp M_z m^*_{\rm W}}{2}\int^{\mu_-}_{\mu_+}{d\epsilon}\epsilon\frac{J_z M_z-2m^*_{\rm W}\epsilon J_\perp M_z}{|\bm d_\epsilon|},
    \end{split}
\end{equation}
and the magnetization of the low-density carriers is: 
\begin{equation}
    m_z = \frac{\rho^*_{\rm W}}{2}\int^{\mu_-}_{\mu_+}{d\epsilon}\frac{J_z M_z-2m^*_{\rm W}\epsilon J_\perp M_z}{2|\bm d_\epsilon|}.
\end{equation}
We conclude observing that the anomalous Hall contribution to the Hall transport~\cite{Haldane_2004} is given by:
 \begin{equation}
     \sigma^{\rm AHE}_{xy}=\frac{e^2}{h}\int^\Lambda\frac{d^2\bk}{2\pi}\Omega_{\bk +} \left[f(\xi_{\bk +})-f(\xi_{\bk -})\right],
 \end{equation}
 where $\Omega_{\bk n}$ is the Berry curvature:
 \begin{equation}
     \Omega_{\bk n}=\frac{1}{k}\partial_k\left(\frac{d_z}{|d|}\right)=\frac{2J^2_\perp J_z \bk^2 \cos^2\theta\sin\theta}{\left[J^2_\perp\bk^4\cos^2\theta+(J_z-J_\perp\bk^2)^2\sin^2\theta\right]^{3/2}}.
 \end{equation}
 We obtain that the anomalous Hall intrinsic contribution to the Hall transport~\cite{Haldane_2004} is proportional to the spin chirality $\bm S_1\cdot(\bm S_2\times\bm S_3)=3\sqrt{3}\cos^2\theta\sin\theta/16$~\cite{supplementary,Binz_2008,verma2022unified,IvarMartin_20108}.

\section{Low-energy theory: including quantum fluctuations}
\label{low_energy_theory}

In this section we derive the low-energy continuum theory describing the fluctuations of the 120$^\circ$ N\'eel order as a continuous medium. The aim is to determine the formation of magnetic polarons where the magnetic order traps single carrier excitations. 

\subsection{Continuum limit: fluctuations of the spin texture}
\label{app:spin_NLSM}

The derivation of the low-energy theory starts from decomposing the 120$^\circ$ order in triangular plaquette with spins $\bm S_j$ oriented along the unit vectors $\bm e_1$, $\bm e_2$ and $\bm e_3$ such that $\bm e_1+\bm e_2+\bm e_3=0$. The latter equality implies that the spins are coplanar and point at angles $120^\circ$ relative to one another. Due to the $z$-axis anisotropy, we fix the classical spin vectors $\bm e_j$ in the $xy$ plane with vectors $\bm e_{1,2,3}=(1,0,0),(-1/2,\sqrt{3}/2,0)$ and $(-1/2,-\sqrt{3}/2,0)$. Following, Ref.~\cite{read1989} we then express the spin Hamiltonian as: 
\begin{equation}
    H_S=\frac{J_H}{6}\sum^{i\,\rm{n.n}\, j}_{i\neq j}\sum_{k=1}^{3}\bm S_{\br_i}\cdot\bm S_{\br_j}^k,
\end{equation}
where $i,j$ sum over the three different sublattices while $k$ are the three primitive directions of the triangular lattice. The continuum limit is taken by performing the gradient expansion: 
\begin{equation}
    \bm S^k_{\br_j}\simeq \bm S_{\br_j}+a(\bm u^k_{ij}\cdot\nabla_{\br})\bm S_{\br_j}+\frac{a^2}{2}(\bm u^k_{ij}\cdot\nabla_{\br})^2\bm S_{\br_j},
\end{equation}
where the vectors $\bm u^k=a(0,1),a(-\sqrt{3}/2,-1/2),a(\sqrt{3}/2,-1/2)$ connects nearest neighboring sites. Summing over $k$ we find: 
\begin{equation}
    H_S=\frac{3J_H}{8}\delta\bm M^2+\frac{3J_Ha^2}{32}\delta\bm M\cdot(\nabla^2_{\br}\delta \bm M)-\frac{9Ja^2}{32}\sum_a\Tr\left[P(R^{-1}\partial_a R)^2\right],
\end{equation}
where we have introduce the net magnetization per plaquette $\delta\bm M=({\bm S}_1+{\bm S}_2+{\bm S}_3)/3$,
$R$ is the SU(2)$_{\rm spin}$ rotation describing a space-dependent twist of the spin texture and $P_{xx}=P_{yy}=1$ while $P_{zz}=0$. As a result of the U(1)$_{\rm spin}$ the relevant fluctuations of the spin textures coupling to the low-energy carriers are the out of plane canting $\delta\bm M=\bm z\sin\theta$ and the twist around the $z$ axis $R=e^{i\phi\sigma^z}$. To lowest order in $\theta$ we have: 
\begin{equation}
    H_S[\theta,\phi]=\frac{3J_H\theta^2}{8}+\frac{9Ja^2}{32}\sum_a(\partial_a\phi)^2.
\end{equation}
Finally, including the Berry connection term we obtain the Lagrangian: 
\begin{equation}
    \mathcal L_S=-\theta\partial_\tau\phi+H_S[\theta,\phi].
\end{equation}
The Kondo coupling $J^{(II)}_K$ between local moments and itinerant carriers gives the contribution: 
\begin{equation}
    H^{(II)}_K =\frac{J_K^{(II)}}{2}\sum_{\br\in \text{W}}c^\dagger_{\br } \bm \sigma c_{\br}\cdot\left(\sum^3_{i=1} \hat{\bm S}_{\br-\bm u_i}\right)=\frac{3J_K^{(II)}}{2}\sum_{\br\in \text{W}}c^\dagger_{\br } \bm \sigma c_{\br}\cdot\delta\bm M_{\br}.
\end{equation}
The Ising SOC locks the spin degree of freedom of the itinerant electrons two the valley $\bm \kappa/\bm \kappa'$. In the long wavelength limit the characteristic variation $|\nabla_{\br}\delta\bm M_{\br}|/|\delta\bm M_{\br}|$ is much smaller than the separation between the two valleys  $|\bm \kappa'-\bm \kappa|$, as a result we find:  
\begin{equation}
    H^{(II)}_K \approx\frac{3J_K^{(II)}}{2\Omega}\int d^2\br\,  \psi^\dagger(\br) \sigma^z \psi(\br) \delta M^z(\br),
\end{equation}
with $\psi=[\psi_\uparrow,\psi_\downarrow]^T$. The double exchange contribution reads: 
\begin{equation}
     H^{(I)}_{K}=\frac{J^{(I)}_K}{2N}\sum_{\bk\bp}\left[V^*_{\bk}V_{\bp}-V_{\bp-\bk}\right]c^\dagger_{\bk}\bm\sigma c_{\bp}\sum_{\br\in{\rm M}} e^{-i(\bk-\bp)\cdot\br} \bm S_{\br}.
\end{equation}
To start with we consider the spin $z$ component. Expanding around $\bm\kappa/\bm\kappa'$ to leading order in momentum we find 
\begin{equation}
     H^{(I)z}_{K}\approx-\frac{3J^{(I)}_K}{2N}\sum_{\bk\bp}^\Lambda c^\dagger_{\bk}\sigma^z c_{\bp}\sum_{\br\in{\rm M}} e^{-i(\bk-\bp)\cdot\br} S^z_{\br}\approx-\frac{3J_K^{(I)}}{2\Omega}\int d^2\br\,  \psi^\dagger(\br) \sigma^z \psi(\br) \delta M^z(\br),
\end{equation}
where the last identity is valid in the long wavelength limit $|\bk-\bp|a\ll1$. Combining the two contributions we find: 
\begin{equation}
    H^z_K=H^{(I)}_K+H^{(II)}_K\approx-\frac{J_z}{2\Omega}\int d^2\br\,  \psi^\dagger(\br) \sigma^z \psi(\br) \delta M^z(\br),
\end{equation}
with $J_z=3\left[J^{(I)}_K-J^{(II)}_K\right]$. For small out of plane deviations of the spin texture $\delta M^z(\br)\approx \theta(\br)/2$ with $\theta(\br)$ canting angle, giving the result: 
\begin{equation}
    H^z_K=-\frac{J_z}{4\Omega}\int d^2\br\,  \psi^\dagger(\br) \sigma^z \psi(\br) \theta(\br).
\end{equation}
The in plane contribution reads: 
\begin{equation}
     H^{(I)\perp}_{K}=\frac{J^{(I)}_K}{2N}\sum_{\bk\bp}\left[V^*_{\bk}V_{\bp}-V_{\bp-\bk}\right]c^\dagger_{\bk\uparrow} c_{\bp\downarrow}\sum_{\br\in{\rm M}} e^{-i(\bk-\bp)\cdot\br} S^-_{\br}+h.c..
\end{equation}
Taking the continuum limit $c_{\bk\uparrow}=c_{\bk+\bm \kappa\uparrow}\approx\psi_{\bk\Uparrow}$ and $c_{\bk\downarrow}=c_{\bk+\bm \kappa'\downarrow}\approx\psi_{\bk\Downarrow}$ with $|\bk|a\ll1$ that is under control in the low-density regime we find
\begin{equation}
     H^{(I)\perp}_{K}\approx\frac{J^{(I)}_K}{2N}\sum^\Lambda_{\bk \bm q}\left[V^*_{\bm\kappa+\bk}V_{\bm\kappa'+\bk+\bm q}-V_{\bm Q+\bm q}\right]\psi^\dagger_{\bk\Uparrow} \psi_{\bk+\bm q\Downarrow}\sum_{\br\in{\rm M}} e^{i(\bm Q+\bm q)\cdot\br} S^-_{\br}+h.c.,
\end{equation}
where $\bm Q=\bm\kappa'-\bm\kappa$ is the momentum modulation connecting the two valleys. Decomposing the sum over $\br$ in the $\sqrt{3}\times\sqrt{3}$ enlarged unit cell with primitive vectors $\tilde{\bm a}_1=\bm \gamma_1-\bm \gamma_2$ and $\tilde{\bm a}_2=\bm \gamma_1-\bm \gamma_3$. By definition $\bm Q\cdot\tilde{\bm a}_{1/2}=2\pi$, as a result we find: 
\begin{equation}
    \sum_{\br\in\Lambda}\sum_{j=1}^3 e^{i\bm q\cdot(\br +\bm d_j)}\omega^{j-1} S^-_{j\br}\approx \sum_{\br\in\Lambda}e^{i\bm q\cdot\br}\sum_{j=1}^3 \omega^{j-1} S^-_{j\br},
\end{equation}
where in the last passage we employed $|\bm q|a\ll1$. Thus, to leading order in $\theta$ and in the in plane twist angle $\phi$ we obtain: 
\begin{equation}
\sum_{\br\in\Lambda}e^{i\bm q\cdot\br}\sum_{j=1}^3 \omega^{j-1} S^-_{j\br}\approx \frac{N}{2}\delta_{\bm q,0}-i\frac{\phi_{\bm q}}{2},
\end{equation}
where $\phi_{\bm q}$ is the Fourier transform of the in plane rotation angle around the $z$ axis. Taking the continuum limit we have: 
\begin{equation}
    \begin{split}
    H^{(I)\perp}_{K}=&\frac{J_\perp}{4}\int \frac{d^2\br}{\Omega} \psi^\dagger_{\Uparrow}(\br)(\hat k_x-i\hat k_y)^2\psi_{\Downarrow}(\br)+h.c.\\
    &+\frac{J_\perp}{2\sqrt{3}}\int \frac{d^2\br}{\Omega}\bm \psi^\dagger\sigma^x\bm\psi\,\partial_y\phi(\br)+\bm \psi^\dagger\sigma^y\bm\psi\,\partial_x\phi(\br),
    \end{split}
\end{equation}
where $\hat k_a=-i\partial_a$ and $\phi$ is the in plane angle describing smooth space dependent rotation of the 120$^\circ$ order around $S^z$. We conclude that in the continuum limit the effective Hamiltonian describing the coupling between the itinerant carriers and the local moments reads: 
\begin{equation}
    \begin{split}
    H_K&=-\frac{J_z}{4}\int \frac{d^2\br}{\Omega}\,  \psi^\dagger(\br) \sigma^z \psi(\br) \theta(\br)+\frac{J_\perp}{4}\int \frac{d^2\br}{\Omega} \psi^\dagger_{\Uparrow}(\br)(\hat k_x-i\hat k_y)^2\psi_{\Downarrow}(\br)+h.c.\\
    &+\frac{J_\perp}{2\sqrt{3}}\int \frac{d^2\br}{\Omega}\left[\bm \psi^\dagger\sigma^x\bm\psi\,\partial_y\phi(\br)+\bm \psi^\dagger\sigma^y\bm\psi\,\partial_x\phi(\br)\right].
    \end{split}
\end{equation}
Additionally, we have the quantum dynamics of the fluctuations of the 120$^\circ$ ordered state. Considering only the collective modes coupled to the itinerant carriers we find: 
\begin{equation}
    \mathcal S_s=\int \frac{d^2\br}{\Omega} dt\left[\theta\partial_t\phi-\frac{3
J_H}{8}\theta^2-\frac{3J_H}{32}\left(\nabla_{\br}\phi\right)^2\right].
\end{equation}

\subsubsection{Variational solution}
\label{app:variational}

In the following we consider a variational ansatz to determine the stability of the spin polaron. To this aim we minimize the Lagrangian $\mathcal L$ given in the manuscript with respect to $\bm \psi$, $\theta$ and $\bm e$. For the sake of simplicity, we limit our analysis to the simple case $J_\perp=0$ and the variational wavefunction simply reduces to:
\begin{equation}
    \bm\psi(r,\phi)=\frac{1}{\sqrt{\pi}\lambda}e^{-r^2/2\lambda},
\end{equation}
where $p\in[0,1]$ and by definition $\int_{\br}|\bm\psi|^2=1$. We readily find: 
\begin{equation}
    E_{\bm \psi}=\int_{\br} \left[\bm\psi^\dagger \frac{\bm k^2}{2m^*_{\rm W}}\bm\psi-\frac{J^2_z}{24 J_H}|\psi|^4\right]=\frac{1}{2m^*_{\rm W}\lambda^2}-\frac{J^2_z}{48 \pi J_H\lambda^2}.
\end{equation}
The variational energy scales as $\lambda^{-2}$ implying that there are two different solutions: for $E_{\bm\psi}>0$ the delocalized solution with $\lambda=\infty$. On the other hand, for $E_{\bm\psi}<0$ we have a polaron state and due to the scaling $\sim\lambda^{-2}$ the polaron collapses into a point $\lambda=0$. The critical line defining the region where the polaron bound state takes place is obtained by solving $F_{\bm\psi}=\lambda^2 E_{\bm\psi}=0$  leading to $J^c_z=12\sqrt{\pi J_H t_{\rm W}/2}$.

\section{Impurity Hamiltonian and one loop renormalization group equations}

In this section we develop the simplest non-trivial approach to describe the competition between magnetism and the screening of the local moment. Instead of a lattice of local moments we focus on a single isolated site and we derive the renormalization group equations at one loop.

\subsection{The Hamiltonian in the single site limit and poor man's scaling laws}

Considering the single impurity problem we now determine the poor's man scaling phase diagram. Focusing on the $\varphi=2\pi/3$ relevant for AB-stacked heterobilayer the conduction electron Hamiltonian expanded around $K_\uparrow=K$ and $K_\downarrow=K'$ reads
\begin{equation}
    \mathcal H= \sum^\Lambda_k\sum_\alpha \left(\frac{k^2}{2m^*_{\rm W}}-\mu\right)\chi^\dagger_{\alpha}(k)\chi_{\alpha}(k), 
\end{equation}
where $\alpha=(m,\sigma)$ with $m$ eigenvalue of $L_z$. In the latter expression we have expanded the Fermi field operators in partial waves $\Psi_{\bk\sigma}=\sum_{m}e^{im\theta_{\bk}}\chi_{m\sigma}(k)$.
In the single site limit the exchange reads: 
\begin{equation}\label{single_site_hamiltonian}
    H_{K}=\frac{S^z }{2}\sum_{\bk,\bk'}^{\Lambda}\left(J_{\kappa+\bk,\kappa+\bk'} \Psi_{\bk\uparrow}\Psi_{\bk'\uparrow}-J_{\kappa'+\bk,\kappa'+\bk'} \Psi_{\bk\downarrow}\Psi_{\bk'\downarrow}\right)+\frac{S^+}{2}\sum_{\bk,\bk'}^{\Lambda}J_{\kappa'+\bk,\kappa+\bk'} \Psi_{\bk\downarrow}\Psi_{\bk'\uparrow}+h.c.
\end{equation}
To express the exchange in terms of $\chi$ we first expand the couplings for small $|\bm k|a\ll 1$
\begin{equation}\begin{split}
    &J_{\kappa'+\bk,\kappa+\bk'} \approx \frac{3a^2}{4}J^{(I)}_K k_+ k'_++\frac{J_z}{2\sqrt{3}}ia(k-k')_-+\frac{J_z}{24}a^2(k-k')^2_+,\\
    &J_{\kappa+\bk,\kappa+\bk'}\approx-J_z+\frac{3a^2}{4}J^{(I)}_Kk_-k'_+ + \frac{J_z}{12}a^2(\bk-\bk')^2,
\end{split}\end{equation}
leading to the result given in the manuscript.

\subsubsection{Poor man's scaling for the $p$-wave $J_z=0$ channel}

For the $p-$wave channel the one loop correction of the coupling constant can be readily obtained and reads: 
\begin{equation}
    \delta J_{p}=\Delta_p(T) J^2_{p},
\end{equation}
where 
\begin{equation}
\begin{split}
    \Delta_p(T)&=-\frac{T}{N}\sum_{i\epsilon}\sum^\Lambda_q \frac{q^2}{i\epsilon(i\epsilon-\epsilon_q+\mu)}\\
    &=m^*\rho\int^{9t_{\rm W}}_0 d\epsilon \frac{\epsilon}{\epsilon-\mu}\tanh\frac{\epsilon-\mu}{2T}. 
\end{split}
\end{equation}
We find: 
\begin{equation}
\begin{split}
    \Delta_p(T/s)-\Delta_p(T) &=\frac{m^*\rho\log s}{2}\int^{\infty}_{-\mu/T}dx\frac{x+\mu/T}{\cosh^2 x/2}\\
    &\approx 2 m^*\rho \mu \log s,
\end{split}
\end{equation}
for $T\ll \mu$.
Thus, the renormalisation group equation obtained by scaling the temperature of the conduction electrons $T\to T/ s$ with $s>1$ reads:  
\begin{equation}
    \frac{d J_{p}}{d\log s} = m^*_W x J^2_{p},
\end{equation}
where $x=2\rho\mu$ is the number of itinerant carriers per unit cell. Integrating the latter equation we readily find: 
\begin{equation}
    J_p(s)=\frac{J_p(s_0)}{1-x m^*_{\rm W}J_{p}(s_0) \log(s/s_0)}.
\end{equation}
Starting from $J_{p}(s_0)=J_\perp-J_z/12 >0$ we find a log singularity for $s^*=s_0 \exp{1/[m^*_{\rm W} x J_{p}(s_0)]}$, thus, noticing that $s^*/s_0=E_\Lambda/T_K$ we found $T_K=E_\Lambda e^{-1/[m^*_{\rm W} x J_p(s_0)]}$.

\subsubsection{Poor man's scaling for the $s$-wave and the $p-$wave $J_z=\pm2$ channels}

The $s-$wave and $J_z=\pm2$ $p-$wave channels are coupled to the local moment via the 
\begin{equation}
    \begin{split}
    \mathcal H^{sp}_K=-\frac{J_z-\delta J_z}{2} S^z\bm \chi^\dagger_{0}\sigma^z\bm \chi_0-\frac{J_z}{24}S^z \left(\chi^\dagger_{+\uparrow}\chi_{+\uparrow}-\chi^\dagger_{-\downarrow}\chi_{-\downarrow}\right)+\frac{J_z}{4\sqrt{3}} \left( S^+\chi^\dagger_{-\downarrow}\chi_{0\uparrow}+S^+\chi^\dagger_{0\downarrow}\chi_{+\uparrow}+h.c.\right),
    \end{split}
\end{equation}
where we neglected term of the order $k^2$ and $k'^2$ involving $d$-wave angular momenta. The first order correction of the exchange coupling $J^z_s$ reads: 
\begin{equation}
    \delta J^z_s = \Delta_p(T) J_\perp^2. 
\end{equation}
On the other hand, the correction to the $J^z_p$ exchange is given by: 
\begin{equation}
    \delta J^z_p = \Delta_s(T) J_\perp^2, 
\end{equation}
with 
\begin{equation}
\begin{split}
    \Delta_s(T) &= -\frac{T}{N}\sum_{i\epsilon}\sum^\Lambda_q \frac{1}{i\epsilon(i\epsilon-\epsilon_q+\mu)}\\
    &=\frac{\rho}{2}\int^W_0 d\epsilon \frac{1}{\epsilon-\mu}\tanh\frac{\epsilon-\mu}{2T}.
\end{split}
\end{equation}
In this case the variation of $\Delta_s(T)$ to a rescaling of $T$ reads: 
\begin{equation}
\begin{split}
    \Delta_s(T/s)-\Delta_s(T) &=\frac{\rho\log s}{2}\int^{\infty}_{-\mu/T}\frac{dx}{\cosh^2 x/2}\\
    &\approx 2 \rho \log s.
\end{split}
\end{equation}
Finally, we have the correction to the transverse coupling $J^\perp$: 
\begin{equation}
    \delta J^\perp=J^\perp\frac{J^z_s\Delta_s(T)+J^z_p \Delta_p(T)}{2}.
\end{equation}
The scaling laws are given by: 
\begin{equation}
\frac{dJ_\perp}{d\log s}=J_\perp\frac{2\rho J^s_z + x m^*_{\rm W} J^p_z}{2},\,\frac{dJ^p_z}{d\log s}=2\rho{J_\perp}^2,\,\frac{dJ^p_z}{d\log s}= m^*_{\rm W}x {J_\perp}^2.
\end{equation}



\end{document}